\documentclass[aps,pre,twocolumn,showpacs,showkeys,groupedaddress,floatfix]{revtex4-1}
\usepackage[T2A]{fontenc}
\usepackage[cp1251]{inputenc}
\usepackage[english,russian]{babel}
\usepackage{graphicx}
\usepackage{subfigure}
\usepackage{multirow}
\usepackage{amsmath}
\usepackage{amssymb}

\graphicspath{{figures/}}

\begin{document}
\selectlanguage{english}
\title{Random sequential adsorption of partially oriented linear $k$-mers on square lattice}

\author{Nikolai I. Lebovka}
\email{lebovka@gmail.com}

\affiliation{Institute of Biocolloidal Chemistry named after F.D.
Ovcharenko, NAS of Ukraine, 42, blvr. Vernadskogo, 03142 Kiev,
Ukraine}

\author{Natalia N. Karmazina}
\email{tusiatutsi@rambler.ru}

\author{Yuri Yu. Tarasevich}
\email[Corresponding author: ]{tarasevich@aspu.ru}
\affiliation{Astrakhan State University, 20a Tatishchev Str, 414056
Astrakhan, Russia}

\author{Valeri V. Laptev}
\email{serpentvv@mail.ru}

\affiliation{Astrakhan State Technical University, 16 Tatishchev
Str, 414056 Astrakhan, Russia}

\date{\today}

\begin{abstract}
Jamming phenomena on a square lattice are investigated for two
different models of anisotropic random sequential adsorption (RSA) of linear
$k$-mers  (particles occupying $k$ adjacent adsorption sites along a line).
The length of a $k$-mer varies from 2 to 128. Effect of $k$-mer alignment on the jamming threshold is examined. For completely ordered systems where all the $k$-mers are aligned along one direction (\emph{e.g.}, vertical), the obtained simulation data are very close to the known analytical results for 1d systems. In particular, the jamming threshold tends to the R{\'e}nyi's Parking Constant for large $k$. In the other extreme case, when
$k$-mers are fully disordered, our results correspond to the published results for short $k$-mers. It was observed that for partially oriented systems the jamming configurations consist of the blocks of vertically and horizontally oriented $k$-mers ($v$- and $h$-blocks, respectively) and large voids between them. The relative areas of different blocks and voids depend on the order parameter $s$, $k$-mer length and type of the model. For small $k$-mers ($k\leqslant 4$), denser configurations are observed in disordered systems  as compared to those of completely ordered systems. However, longer $k$-mers exhibit the opposite behavior.
\end{abstract}


\maketitle

\section{\label{sec:introduction}Introduction}

In the Random Sequential Adsorption (RSA), objects randomly deposit on a substrate; this process is irreversible, and the newly placed objects cannot overlap or pass through the previously deposited ones~\cite{Evans}.
The final state generated by the irreversible adsorption is a disordered one (known as the jamming state); no more objects can deposit in this state due to the absence of any free space of appropriate size and shape~\cite{Lois2008,Biroli2008,Evans}. The fraction of the total surface occupied by the adsorbed objects, which is called the jamming concentration,
is of central interest for understanding of the RSA processes.

The RSA model studies of the objects with different
shape (\emph{e.g.}, squares, ellipses, or stiff rods (needles)) have shown that jamming concentration depends strongly on the object shape and size (see, \emph{e.g.}, \cite{Evans}).
The RSA model has attracted a great interest as a tool for explanation of the wide class of irreversible phenomena observed in adsorption of chainlike and polymer molecules on homogeneous and heterogeneous surfaces \cite{Loscar2003PRE}.
Jamming of short flexible linear chains on a square lattice has been studied by Becklehimer~\cite{Becklehimer1994} and Wang~\cite{Wang1996}.
Adsorption of semi-flexible chains has been recently investigated using RSA model by Kondrat~\cite{Kondrat2002}. RSA of the binary mixtures of extended objects (linear and bent chains) has been described by Lon\v{c}arevi\'{c} \emph{et al.} \cite{Budinski24}. RSA on a two-dimensional (2d) square lattice has been examined for the stiff and flexible polymer chains simulated by a sequence of lattice points forming needles, T-shaped objects and crosses, as well as flexible linear chains and star-branched chains consisting of three and four arms \cite{Adamczyk2008JCP}.
RSA model is widely used for simulation of thin film formation in the process
of nanoparticle deposition~\cite{Brosilow1991,Talbot1989,Budinski76,Budinski24,Rampf2002}.
The interplay between the jamming and percolation phenomena is discussed in several works (\emph{e.g.}, \cite{Vandewalle,Rampf2002,Adamczyk2008JCP,KondratPre63}).

Important exact result for jamming concentration $p_j$ of $k$-mers in one dimension (1d) random sequential adsorption has been reported
by Krapivsky \emph{et al.}~\cite{Krapivsky2010}
\begin{equation}\label{eq:ue1Djamm}
p_{j} = k
\int\limits_{0}^{\infty}\exp\left(-u-2\sum\limits_{j=1}^{k-1}\frac{1-\exp(-ju)}{j}\right)\,du.
\end{equation}

In particular, the jamming concentration for dimers ($k=2$) placed along
one line is $p_{j}=1-\mathrm{e}^{-2} \approx 0.86466$ and for
trimers ($k=3$) is $p_{j} = 3 D (2) - 3 \mathrm{e}^{-3} D (1)
\approx 0.82365$, where $D(x) = \mathrm{e}^{-x^2}\int_{0}^{x}
\mathrm{e}^{t^2} dt$ is Dawson's integral. For $k \to \infty$,
the jamming threshold tends to R{\'e}nyi's Parking Constant
${p_{j}}\to {c_{R}} \approx 0.7475979202$~\cite{Renyi1958}.

Many efforts have been concentrated on the study of deposition of $k$-mers on a discrete 2d substrate.
Most of previous works have been devoted
to the deposition of randomly oriented
linear $k$-mers on square lattices.
In square lattice models, the number of possible orientations of $k$-mers is restricted to 2.
When orientation of $k$-mer is fixed, the situation is equivalent to the above mentioned case of deposition
on a 1d lattice.

For lattice model, Manna and Svrakic observed that deposited $k$-mers ($1 \leqslant k \leqslant 512$) tend to align parallel to each other~\cite{Manna1991JPhysA}. They form large domains (stacks) with voids ranging from a single site up to the length of $k$-mer. The similar stacking and void formation is observed also for continuous RSA model of deposition of infinitely-thin line segments~\cite{Ziff1990JPhysA,Viot1992PhysicaA}.

It is worth noting that principally different jamming behavior is observed for lattice and continuous models.
For example, for deposition of extremely elongated objects $k \to \infty$ (\emph{i.e.}, in the limit of infinite aspect ratio)
the asymptotic jamming concentration $p_j(\infty)$ is 0 in the off-lattice case \cite{Viot1992PhysicaA}. The value of $p_j$ approaches zero with increase of the aspect ratio $k$ and follows the power law
\begin{equation}\label{eq:power}
p_{j}(k) = p_{j}(\infty)+a/k^\alpha
\end{equation}
with $p_{j}(\infty)=0$ and $a=1/(1+2\sqrt{2})\approx 0.26$  \cite{Viot1992PhysicaA}.

In the discrete case, the asymptotic jamming concentration $p_j(\infty)$ remains finite \cite{Manna1991JPhysA, Bonnier1994PRE, Bonnier1996PRE,KondratPre63}. In the latter case, the presence of finite coverage by the infinite $k$-mers has been interpreted as a consequence of the alignment constraint \cite{Bonnier1994PRE, Bonnier1996PRE}.

 Note that for deposition on a discrete lattice \cite{Manna1991JPhysA, Bonnier1994PRE, Bonnier1996PRE,KondratPre63}, the limiting jamming concentration $p_{j}(\infty)$ depends upon the deposition mechanism. For the conventional RSA model (the vacant lattice site is randomly selected and any unsuccessful attempt of deposition of $k$-mer is rejected) and completely orientationally disordered deposition of linear $k$-mers on square lattices, different Monte Carlo studies have given the estimation $p_{j}(\infty)\approx0.66$~\cite{Bonnier1994PRE, KondratPre63}.  For, so called, "end-on" RSA model~\cite{Bonnier1994PRE} (in this model, once a vacant site has been found, the deposition is (randomly) attempted in all the directions until the segment is adsorbed or rejected), the Monte Carlo calculation has given the noticeably smaller value
$p_{j}(\infty)=0.583\pm 0.010$ \cite{Manna1991JPhysA}.

Different functions have been tested to fit the $p_{j}(k)$ dependence for the deposition of randomly oriented $k$-mers on a two-dimensional square lattice (conventional RSA model)
~\cite{Bonnier1994PRE, Bonnier1996PRE,Vandewalle, KondratPre63}. MC data on jamming concentration $p_{j}(k)$ obtained for line segments of length $2\leqslant k\leqslant 512$ on square lattices of linear size $L\leqslant 4096$ (preserving in all cases the ratio $L/k>8$) have been fitted using the series expansion
\begin{equation}\label{eq:Bonnier}
p_{j}(k) = p_{j}(\infty)+a/k+b/k^2,
\end{equation}
and have given $p_{j}(\infty)=0.660\pm 0.002$, $a \approx 0.83$ and $b \approx -0.70$~\cite{Bonnier1994PRE, Bonnier1996PRE}.

The similar data for line segments of length $2 \leqslant k \leqslant 24$ on lattices of linear size $L=2000$ have been approximated using an empirical law~\cite{Vandewalle}
\begin{equation}\label{eq:Vandewalle}
p_{j}(k) = p^*(1-\gamma(1-1/k)^2).
\end{equation}

However, this equation is in fact a particular case of Eq.~\ref{eq:Bonnier} under constrain of $b=-a/2$. It can be easily checked by using the substitutions of $p^*=p_{j}(\infty)+a/2$ and $\gamma=a/(2p^*)$. Fitting of the numerical data presented in Table~\ref{tab:pjvsSk} of~\cite{Vandewalle}) with Eq.~\ref{eq:Bonnier} gives the following estimations: $p_{j}(\infty)=0.684 \pm 0.003$, $a=0.59 \pm 0.01$, $\rho=0.9998$ (coefficient of determination).

Kondrat and P\c{e}kalski have reported the results for $k$-mers with length within $1\leqslant k \leqslant 2000$, lattice size $L=30,100,300,1000,2500$ and more than 100 independent runs~\cite{KondratPre63}. Application of the power law (Eq.~\ref{eq:power}) have given $p_{j}(\infty)=0.66\pm 0.01$, $a \approx 0.44$ and $b \approx 0.77$.

To our best knowledge, the very limited number of works have been
devoted to jamming of non-randomly oriented linear $k$-mers
and only particular case ($k=2$) has been taken into consideration ~\cite{deOliveiraPRA1992, Cherkasova}.
For isotropic problem, the jamming concentration at $k=2$ is $p_j\approx0.9068$ ~\cite{NordJPC1985, deOliveiraPRA1992, Cherkasova}.

In anisotropic problem, the vertical and horizontal orientations occur with different probabilities and degree of anisotropy can be characterized by the order parameter $s$ defined as
\begin{equation}\label{eq:S}
    s =  \left|\frac{N_| - N_-}{N_| + N_-}\right|,
\end{equation}
where $N_|$ and $N_-$ are the numbers of line segments oriented in the vertical and horizontal directions, respectively.

Data of Monte Carlo simulations evidence that the value of $p_j$ decreases with order parameter $s$ increase~\cite{deOliveiraPRA1992,Cherkasova}. In particular case of complete ordering, \emph{i.e.}, at $s=1$, the problem becomes one-dimensional. An interesting finding is that in the limit of $s\to 1$ the asymptotic fraction of dimers with horizontal direction does not vanish but equals to $ \mathrm{e}^{-2}[1-\exp(-2\mathrm{e}^{-2})]/2 \approx 0.016046$~\cite{deOliveiraPRA1992}.

The main goal of the present study is to investigate the effects of $k$-mer length, alignment, and the deposition rules on the
jamming threshold. This investigation is the natural development of the recent work devoted to the dimers, $k=2$ \cite{Cherkasova}. This
work discusses the jamming phenomena~\cite{Evans} for two different kinds of anisotropic sequential deposition of linear $k$-mers on a square
lattice (particles occupying $k$ adjacent adsorption sites along a line).

The paper is organized as follows. Section~\ref{sec:model} contains the basis of the two models of deposition of $k$-mers on a square
lattice. The results obtained using finite size scaling theory and dependencies of the jamming threshold, $p$, and mean radius of pores, $r$, \emph{vs.} order parameter, $s$, are examined and discussed in details in  Section~\ref{sec:results}. We discuss the dependence of the jamming threshold on the model parameters of interest in this Section, too.

\begin{figure}
\centering
\includegraphics[width=\linewidth]{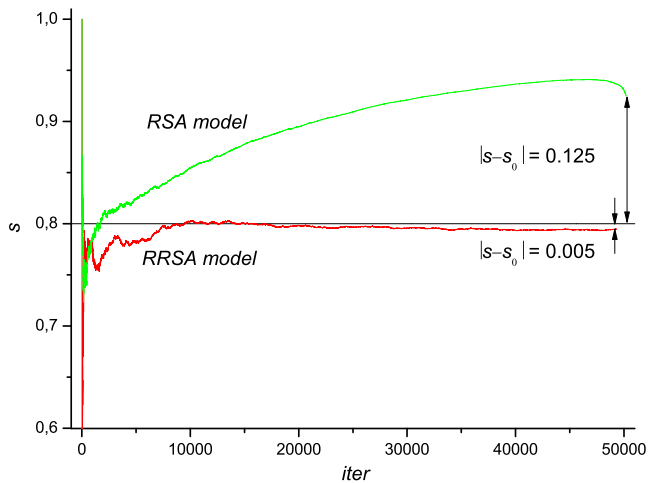}
\caption{Actual order parameter $s$, after each successful attempt to place a new $k$-mer ($k=3$) for RSA and
RRSA models. Lattice size is $L=729$ and predetermined parameter is $s = 0.8$. \label{fig:s_for_iter}}
\end{figure}

\begin{figure*}
\centering
\subfigure[RSA \& RRSA, $s=0.0$]{\includegraphics*[width=0.3\linewidth]{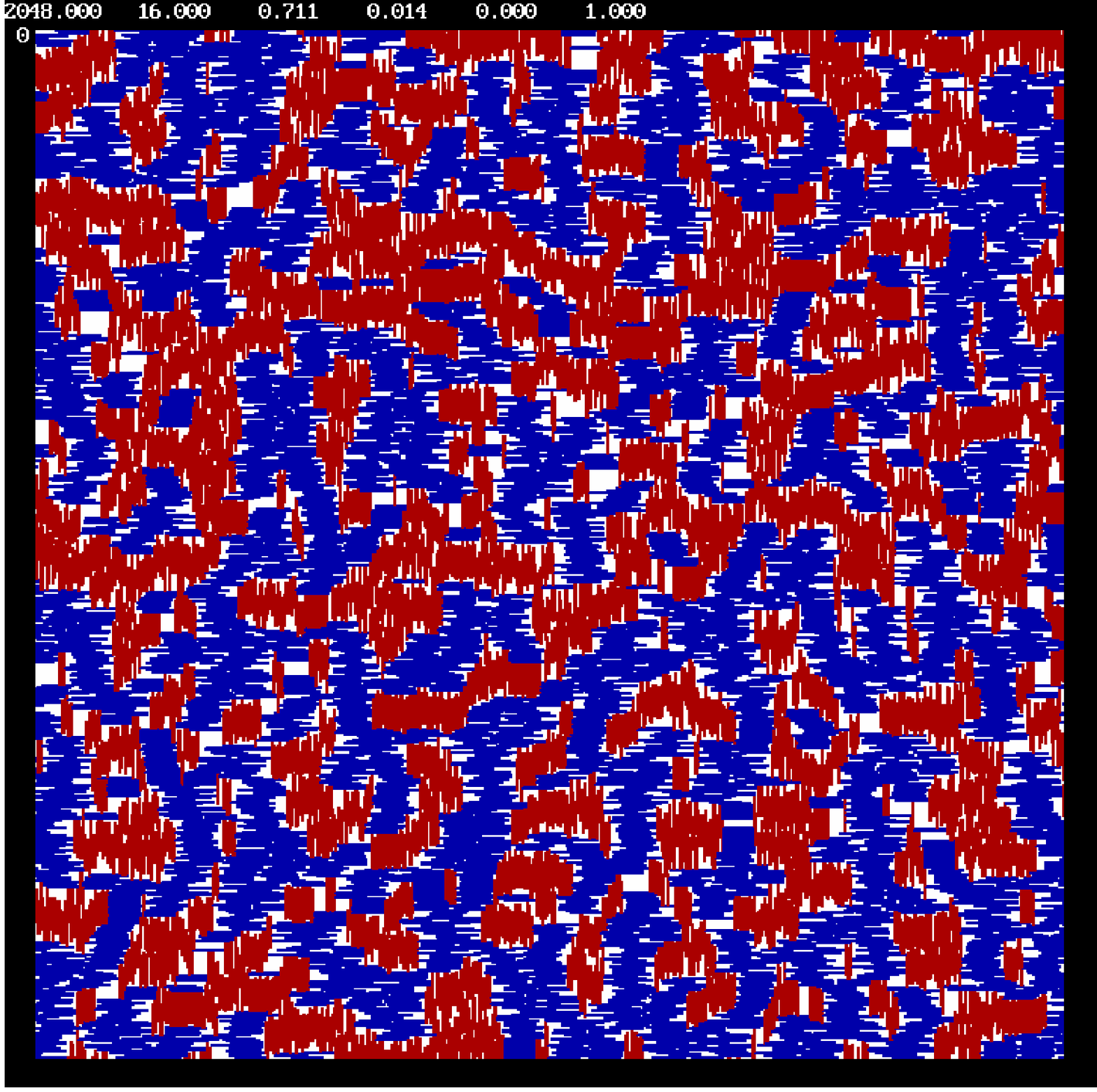}}\hfill
\subfigure[RSA, $s=0.2$]{\includegraphics*[width=0.3\linewidth]{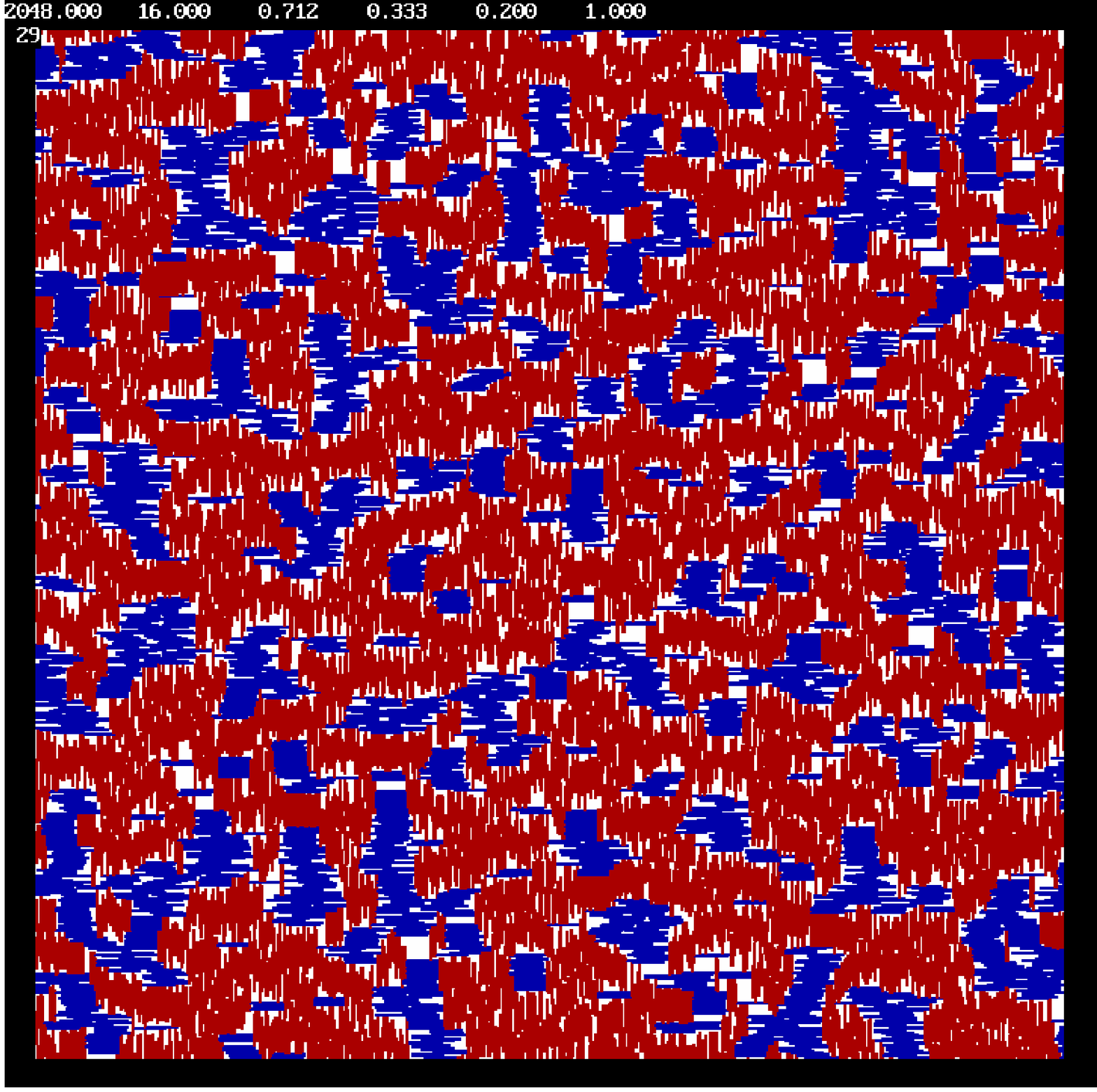}}\hfill
\subfigure[RSA, $s=0.6$]{\includegraphics*[width=0.3\linewidth]{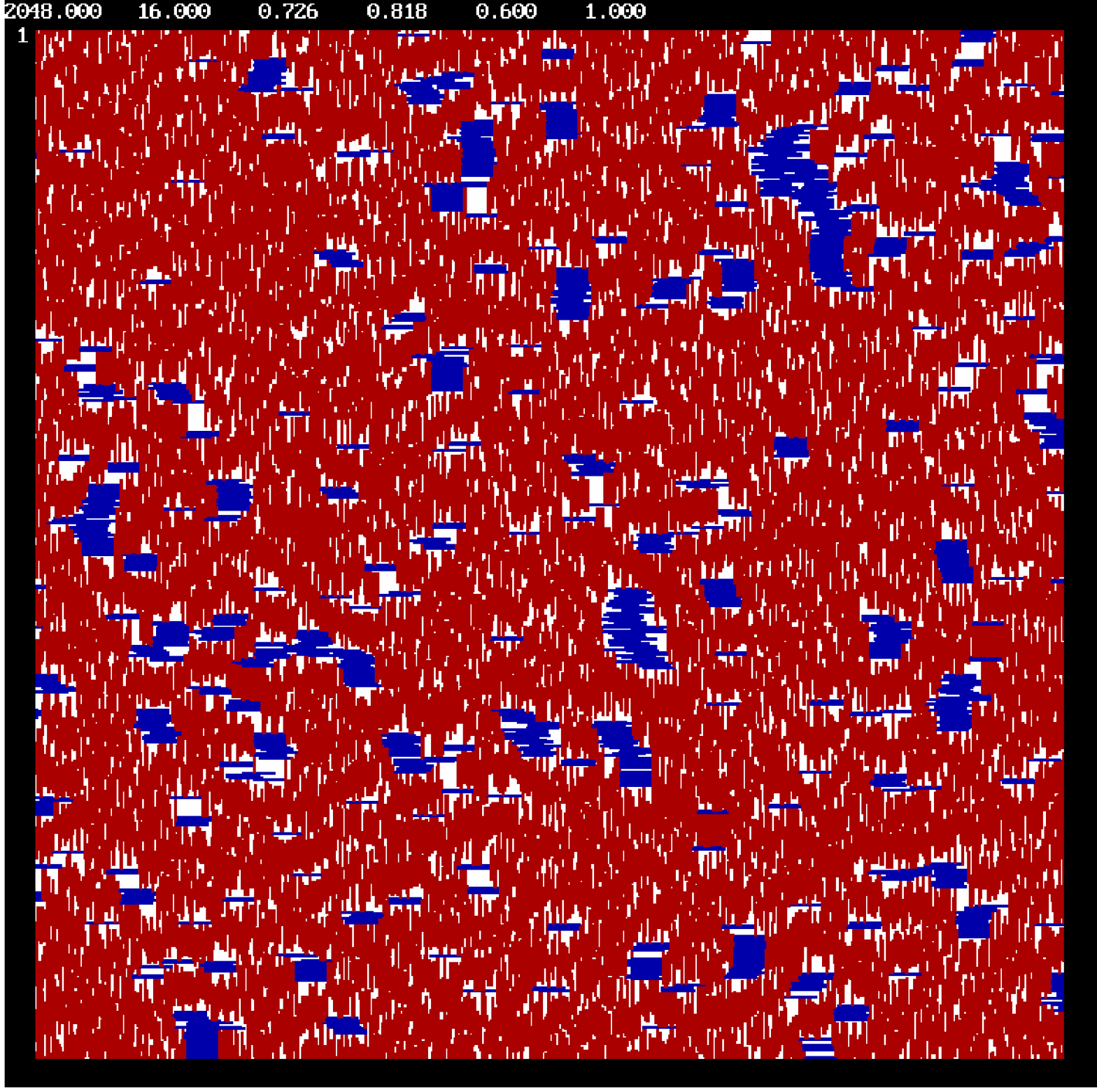}}\\
\subfigure[RSA \& RRSA, $s=1.0$]{\includegraphics*[width=0.3\linewidth]{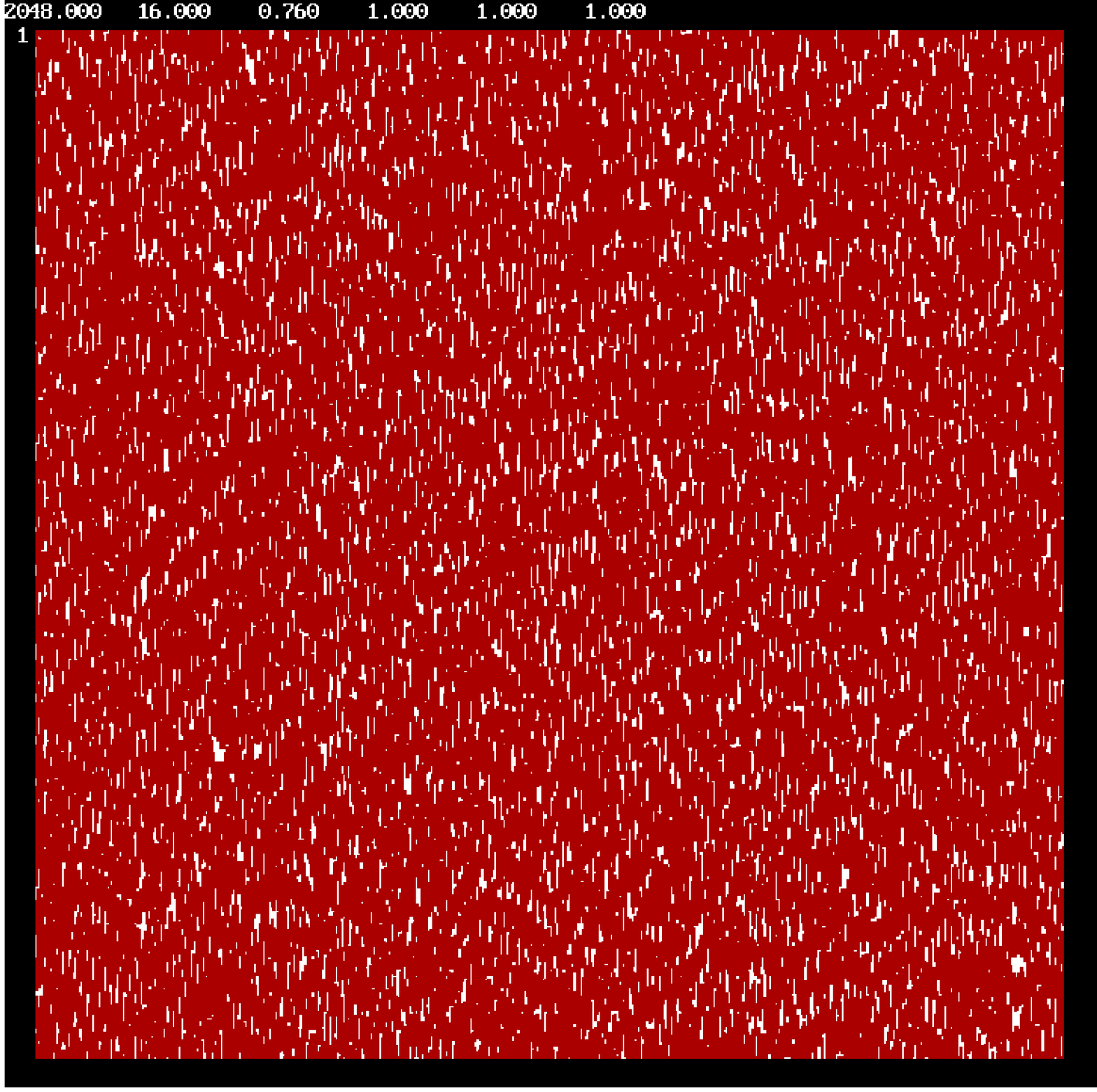}}\hfill
\subfigure[RRSA, $s=0.2$]{\includegraphics*[width=0.3\linewidth]{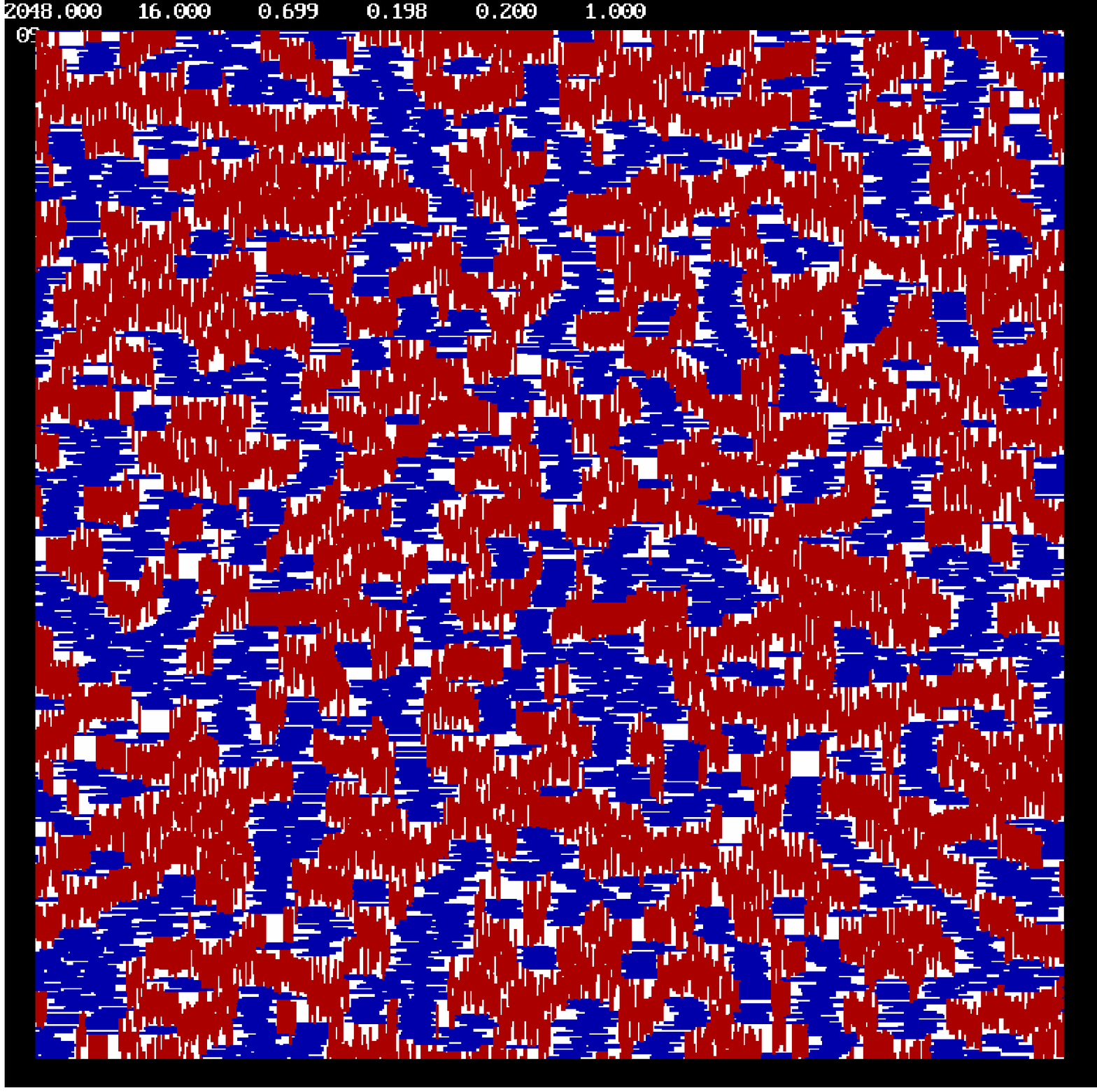}}\hfill
\subfigure[RRSA, $s=0.6$]{\includegraphics*[width=0.3\linewidth]{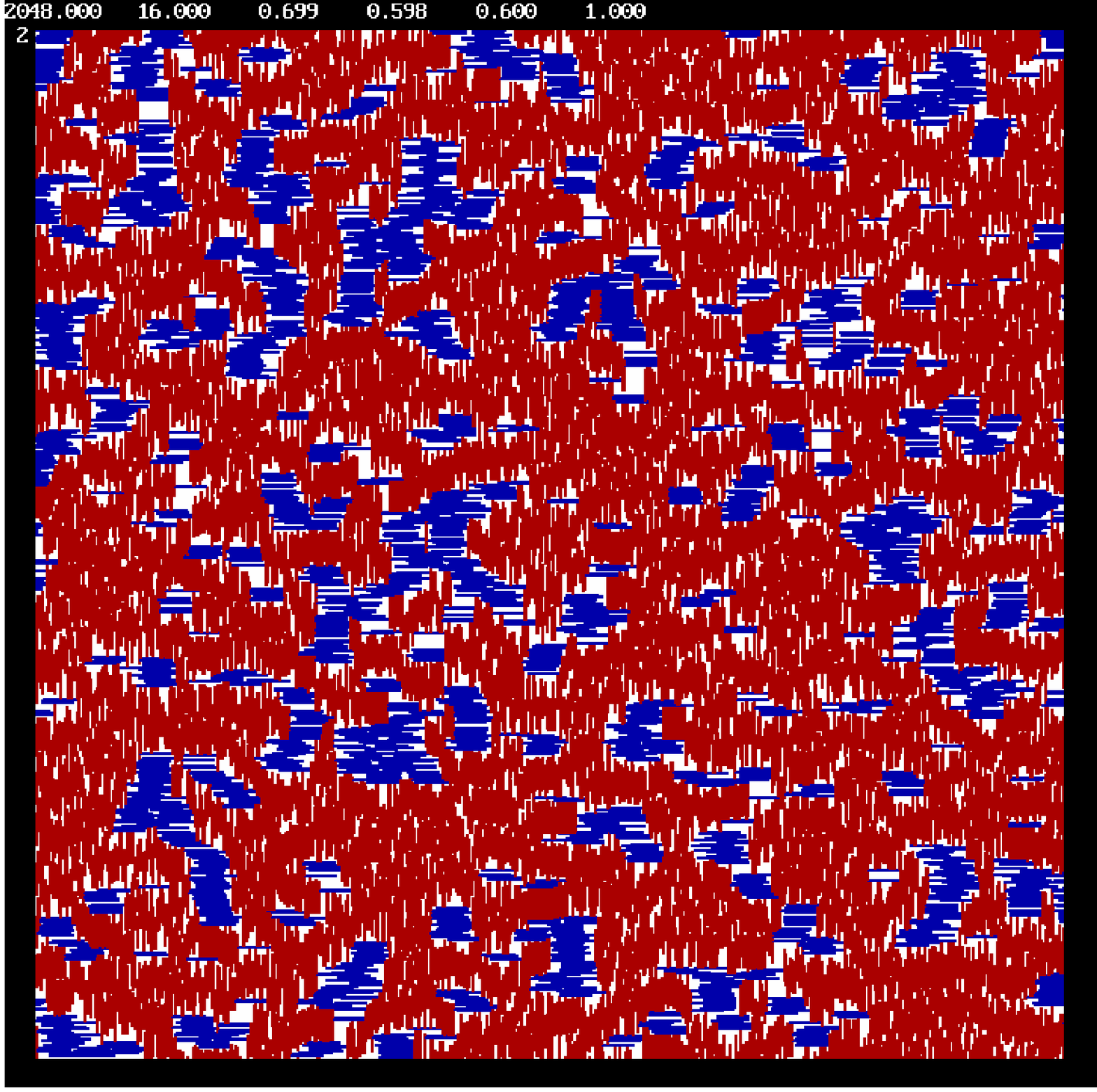}}
\caption{\label{fig:djm_m2}Examples of jamming configurations of $k$-mers ($k=16$) on a square lattice of size $L=2048$ at
different values of order parameter $s$. Vertical and horizontal orientations are represented by different grey levels in printed version and in red in blue in on-line version,  empty sites are labeled white. Here, the part of the lattice sized $512 \times 512$ is shown.}
\end{figure*}

\begin{figure*}
\centering
\subfigure[RSA, $s=0.050$]{\includegraphics*[width=0.3\linewidth]{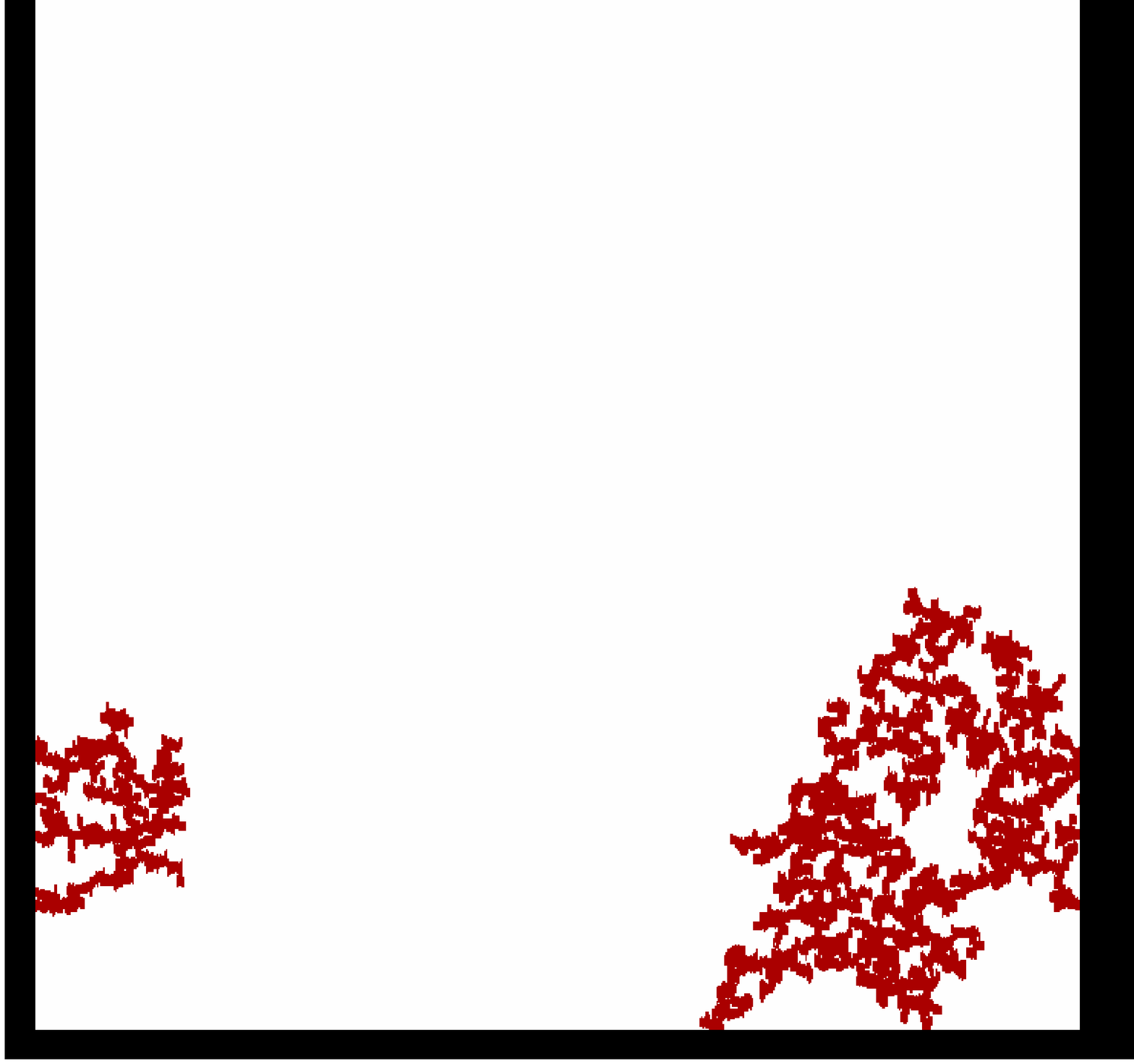}}\hfill
\subfigure[RSA, $s=0.120$]{\includegraphics*[width=0.3\linewidth]{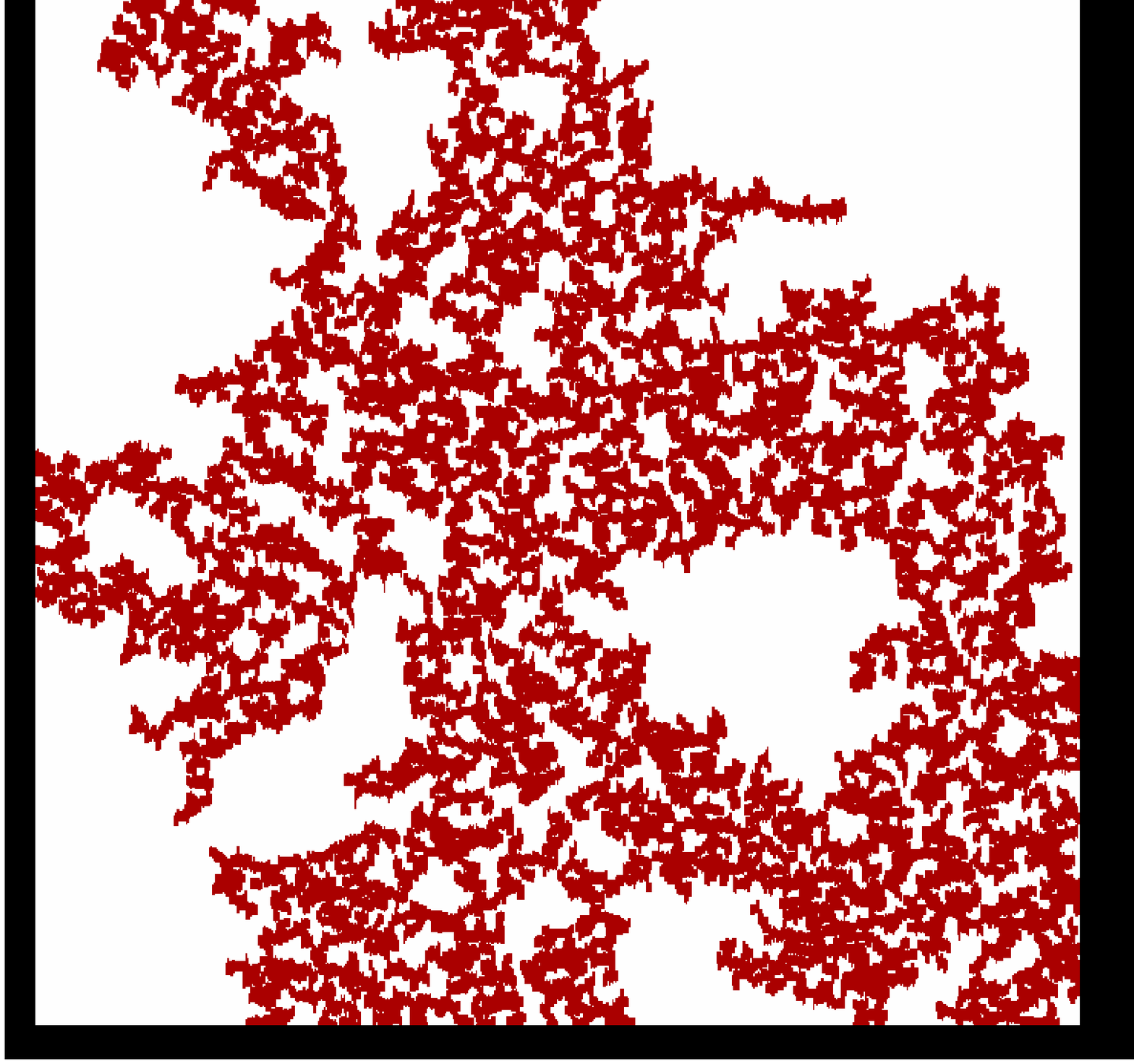}}\hfill
\subfigure[RSA, $s=0.245$]{\includegraphics*[width=0.3\linewidth]{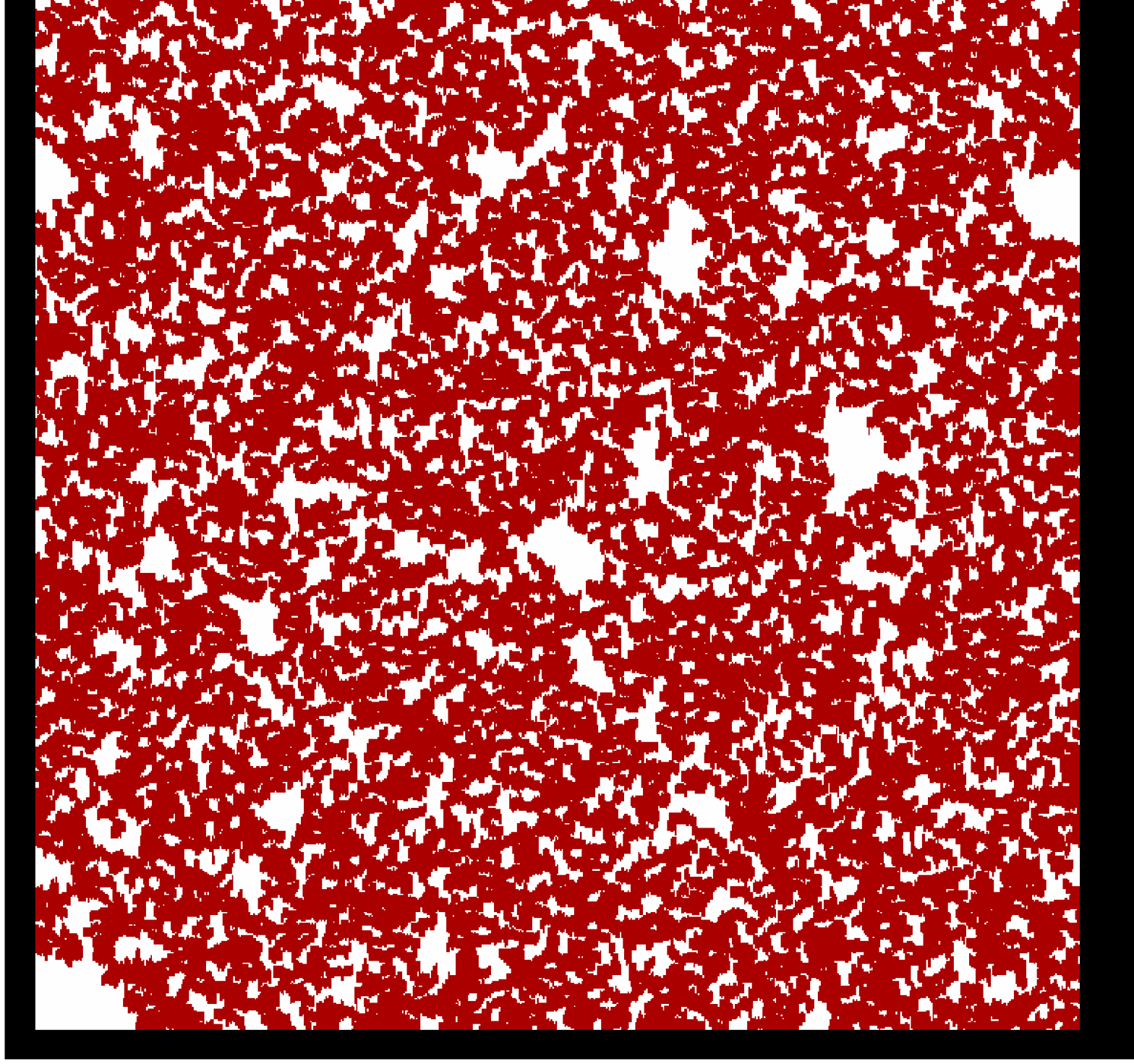}}\\
\subfigure[RRSA, $s=0.050$]{\includegraphics*[width=0.3\linewidth]{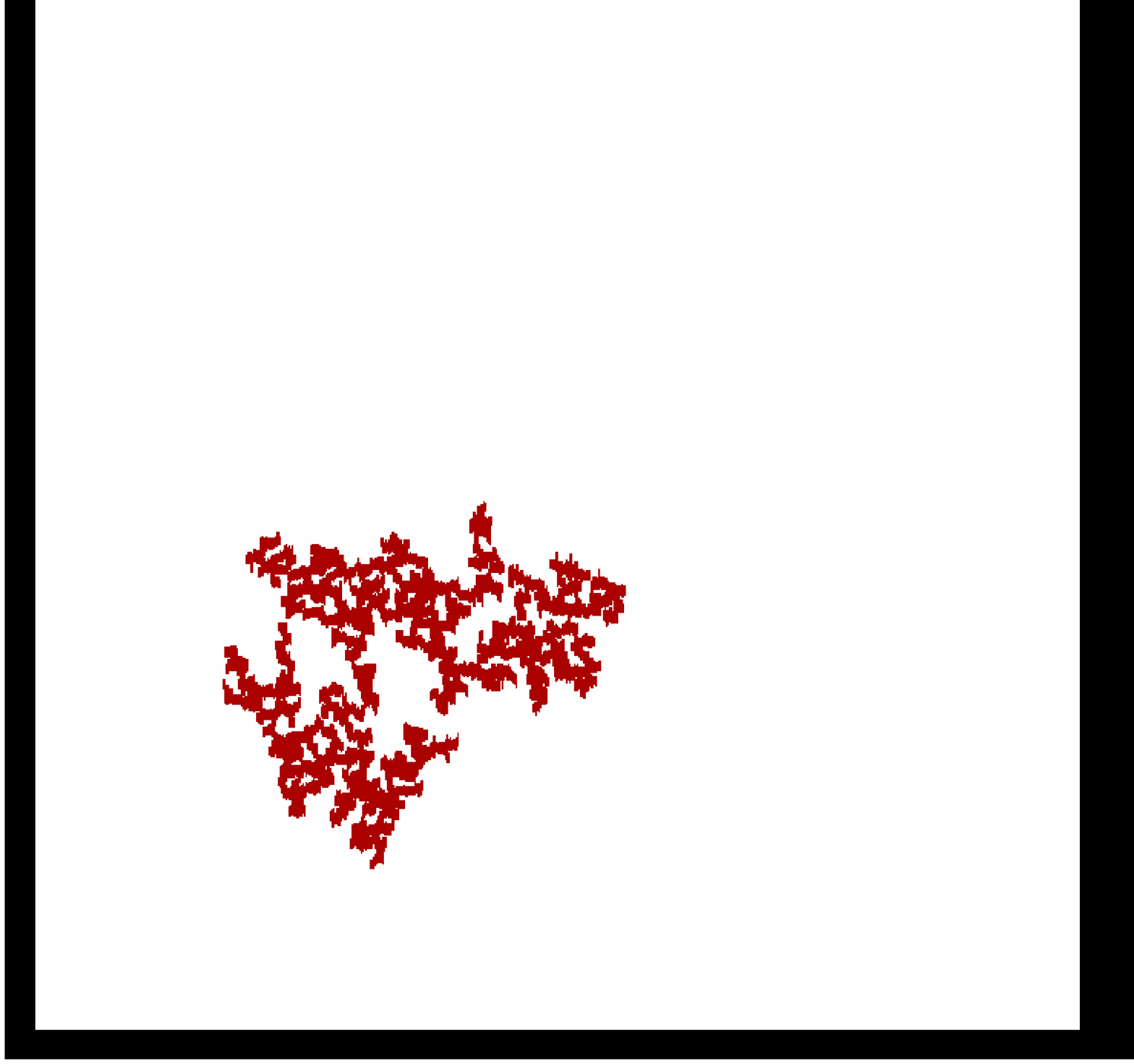}}\hfill
\subfigure[RRSA, $s=0.120$]{\includegraphics*[width=0.3\linewidth]{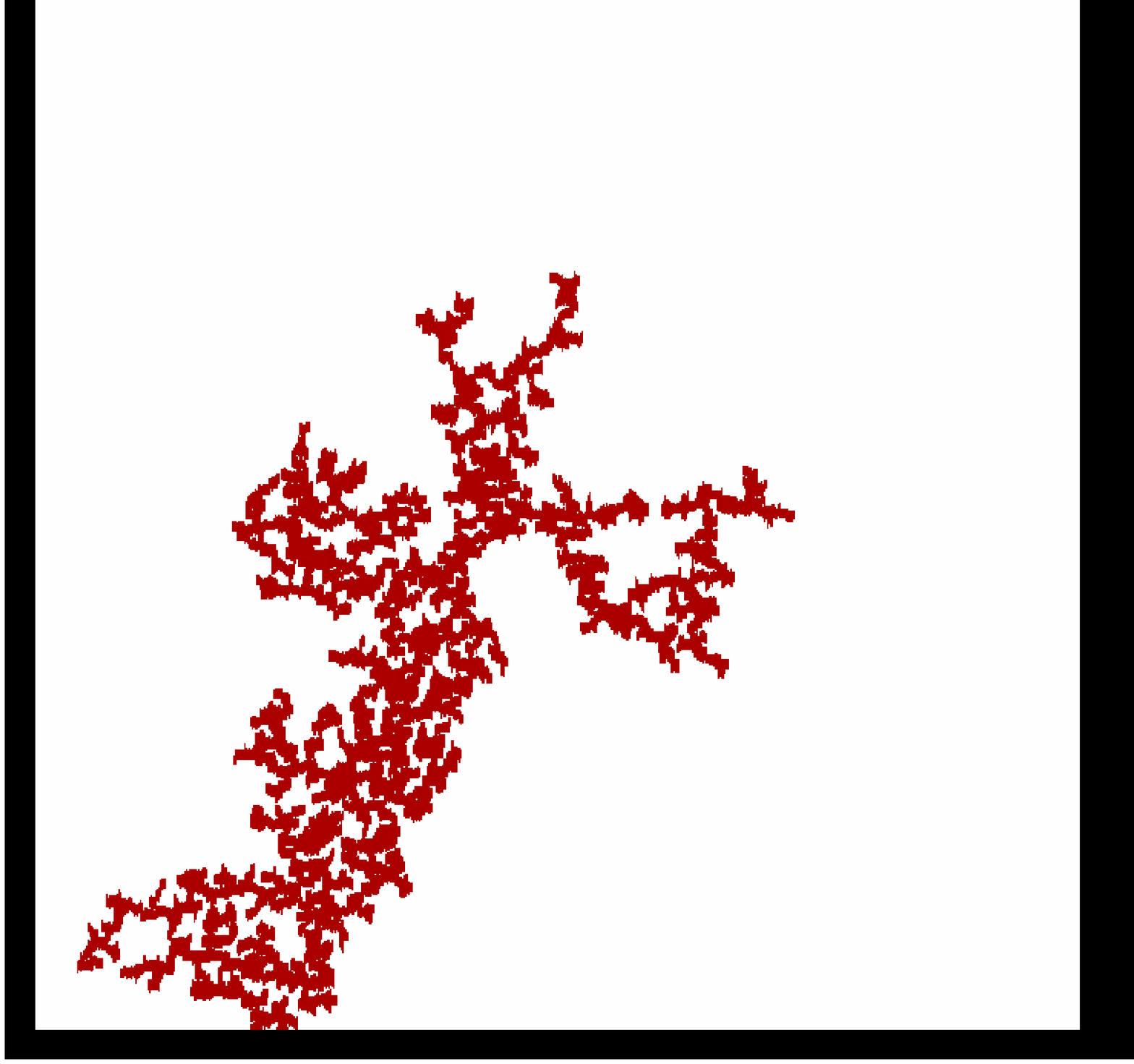}}\hfill
\subfigure[RRSA, $s=0.245$]{\includegraphics*[width=0.3\linewidth]{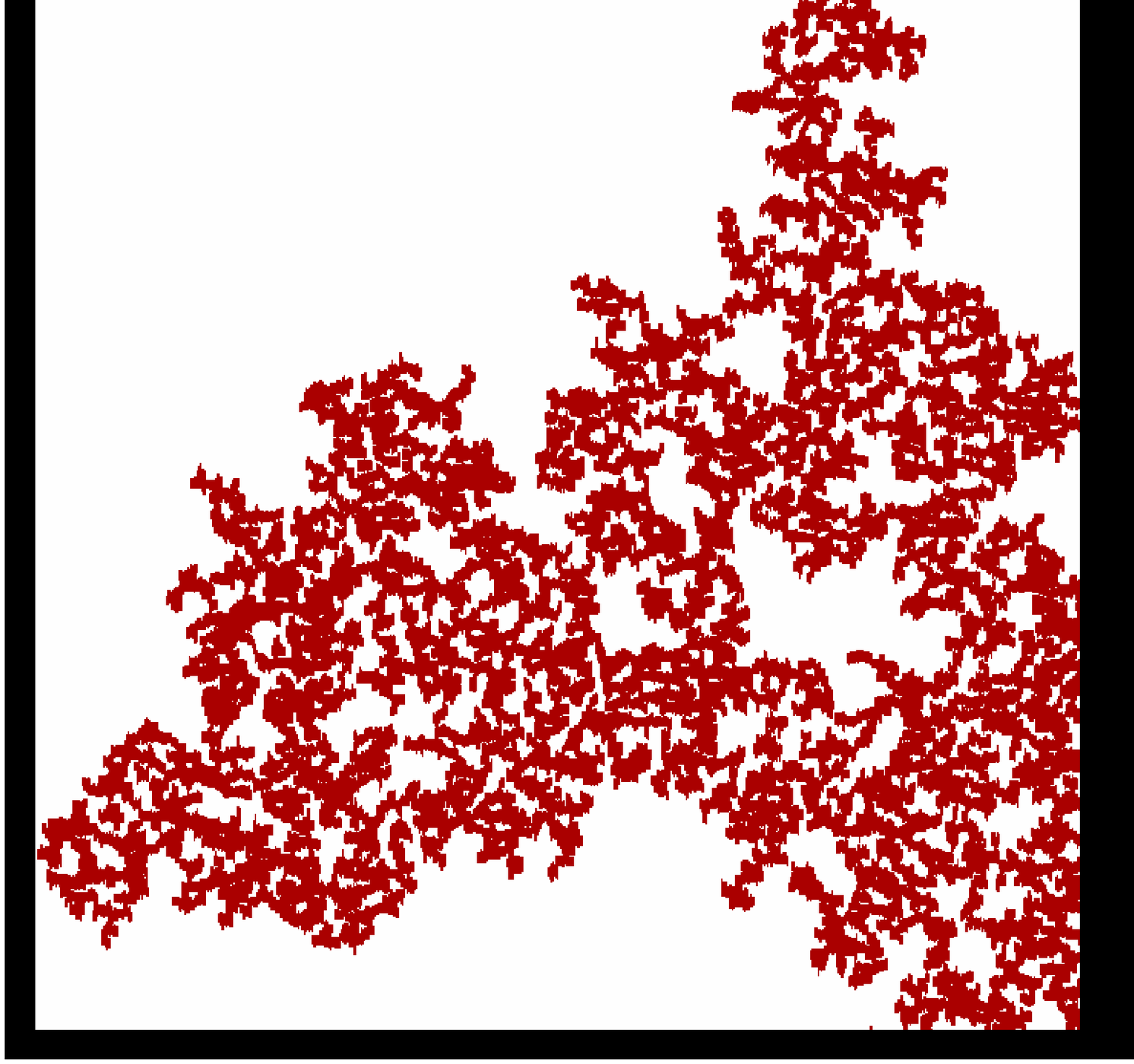}}
\caption{\label{fig:djm_p2}Examples of largest clusters of vertically oriented $k$-mers ($k=16$) for jamming configurations on a square lattice of size $L=2048$ at different values of the order parameter $s$.}
\end{figure*}

\section{\label{sec:model}Description of models and details of simulations }

One can imagine suspension of linear line segments of length $k$ ($k$-mers) in the bulk volume under substrate. In our simulations, a square lattice of linear size $L$ has been used as a substrate, and periodical boundary conditions have been applied both in horizontal and vertical directions (toroidal boundary conditions). The $k$-mers repulse each other and deposit one by one onto a substrate. The anisotropy of $k$-mer orientation in the suspension is predetermined and can be characterized by the input parameter $s$ defined as in Eq.~\ref{eq:S}.

The number of species in the suspension is supposed to be infinitely large, thus deposition doesn't change the anisotropy of the suspension. If different orientations of the deposited objects are not of equal probability (\emph{i.e.}, $N_| \neq N_-$ ), the definition of the jamming state is to be refined~\cite{Cherkasova}. Let us assume $N_| > N_-$. We define jamming for the fixed parameter $s$ as a situation when there exists no possibility of deposition for any additional vertically oriented objects. Nevertheless, there may be places for horizontally oriented objects.

Two different deposition models have been studied. The first is the conventional Random Sequential Adsorption (RSA) model. In this model, the lattice site is randomly selected and an attempt of deposition of a $k$-mer with orientation defined by order parameter $s$ is done. Any unsuccessful attempt is rejected and $k$-mer with a new orientation is selected. The second model, called by us Relaxation Random Sequential Adsorption (RRSA), is very similar to the RSA, however, the unsuccessful attempt is not rejected and a new lattice site is randomly selected until the object will be deposited. Note that in contrast to the known RSA model with diffusion (see, \emph{e.g.}, \cite{Gan1997PRE}), the RRSA model does not restrict the movement of species by the nearest sites only. The species may move all over the substrate searching for a sufficiently large empty space. In both models, the deposition terminates when a jamming state is reached along one of direction.

Physically, the difference between RSA and RRSA models can reflect different binding of $k$-mers near the adsorbing substrate. In RSA model, binding of a $k$-mer by the substrate is weak and the $k$-mer returns to the bulk suspension after an unsuccessful attempt to precipitate. In RRSA model, binding of a $k$-mer to the adsorbing plane is strong, and $k$-mer has an additional possibility of joining the surface after an unsuccessful attempt.

The differences between RSA and RRSA models can be evidently demonstrated by analysis of anisotropy of the deposited layer actually obtained in the course of adsorption (Fig.~\ref{fig:s_for_iter}). The preliminary study has shown that RSA model does not allow preservation of the order parameter $s$. In this model, the substrate "selects" the $k$-mer with appropriate orientation, and it results in deviation of predetermined order parameter $s$ from the actually obtained one, $s_0$. The MC simulation evidences that the value of $s_0$ noticeably exceeds the value of $s$.  On the contrary, RRSA model better preserves the predetermined anisotropy, and $s_0 \approx s$ (Fig.~\ref{fig:s_for_iter}).

\begin{figure}
\centering
\includegraphics[width=\linewidth]{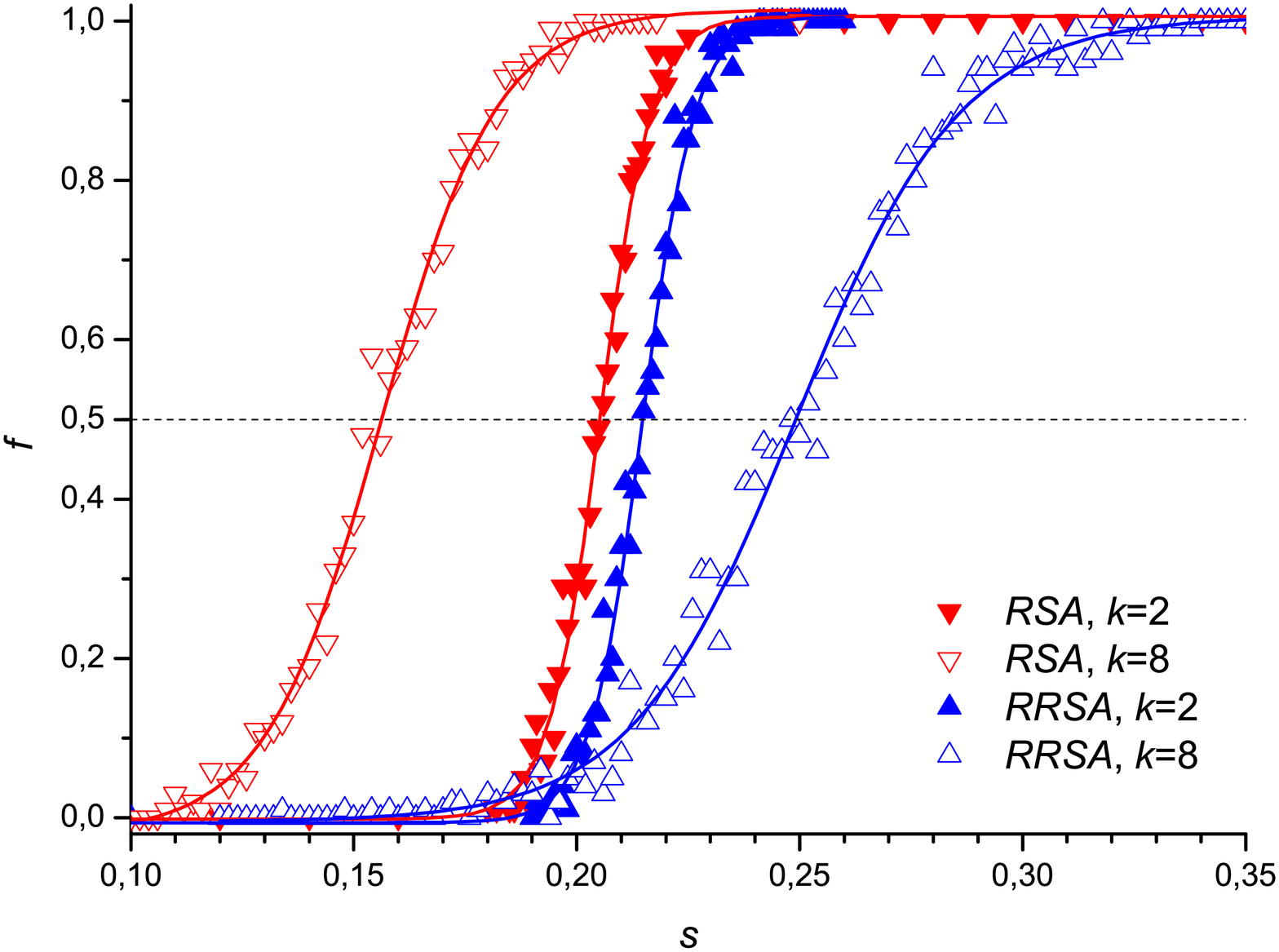}
\caption{Percolation probability $f$ \emph{vs.} predetermined order parameter $s$ for RSA and RRSA models at different values of $k$. Lattice size is $L=1024$ and results are averaged over 100 independent runs. \label{fig:f vs s}}
\end{figure}

\begin{figure}
\centering
\includegraphics[width=\linewidth]{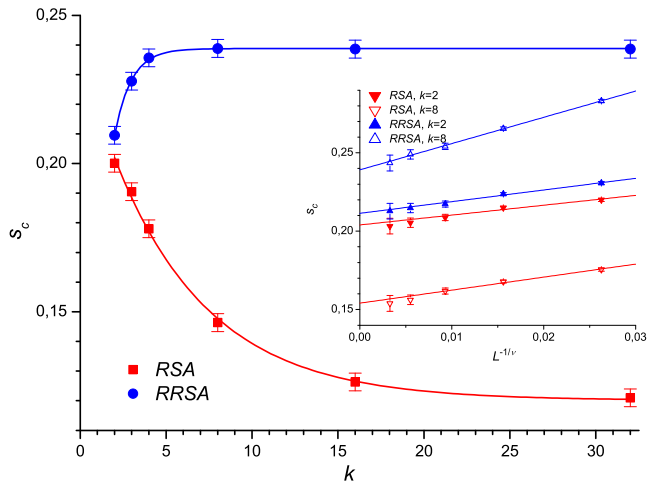}
\caption{Threshold order parameter $s_c=s_c(L \to \infty)$ \emph{vs.} length of $k$-mer for RSA and RRSA models.
Inset shows examples of scaling dependencies in the form of $s_c$ \emph{vs.} $L^{-1/\nu}$  for $k=2$ and $k=8$.
Here, $\nu = 4/3$ is the critical exponent of the correlation length for 2d random percolation problem~\cite{Stauffer}.
 \label{fig:sc_vs k}}
\end{figure}

\begin{figure*}
\centering
\subfigure[$k=3$ (RSA model)]{\includegraphics[width=0.5\linewidth]{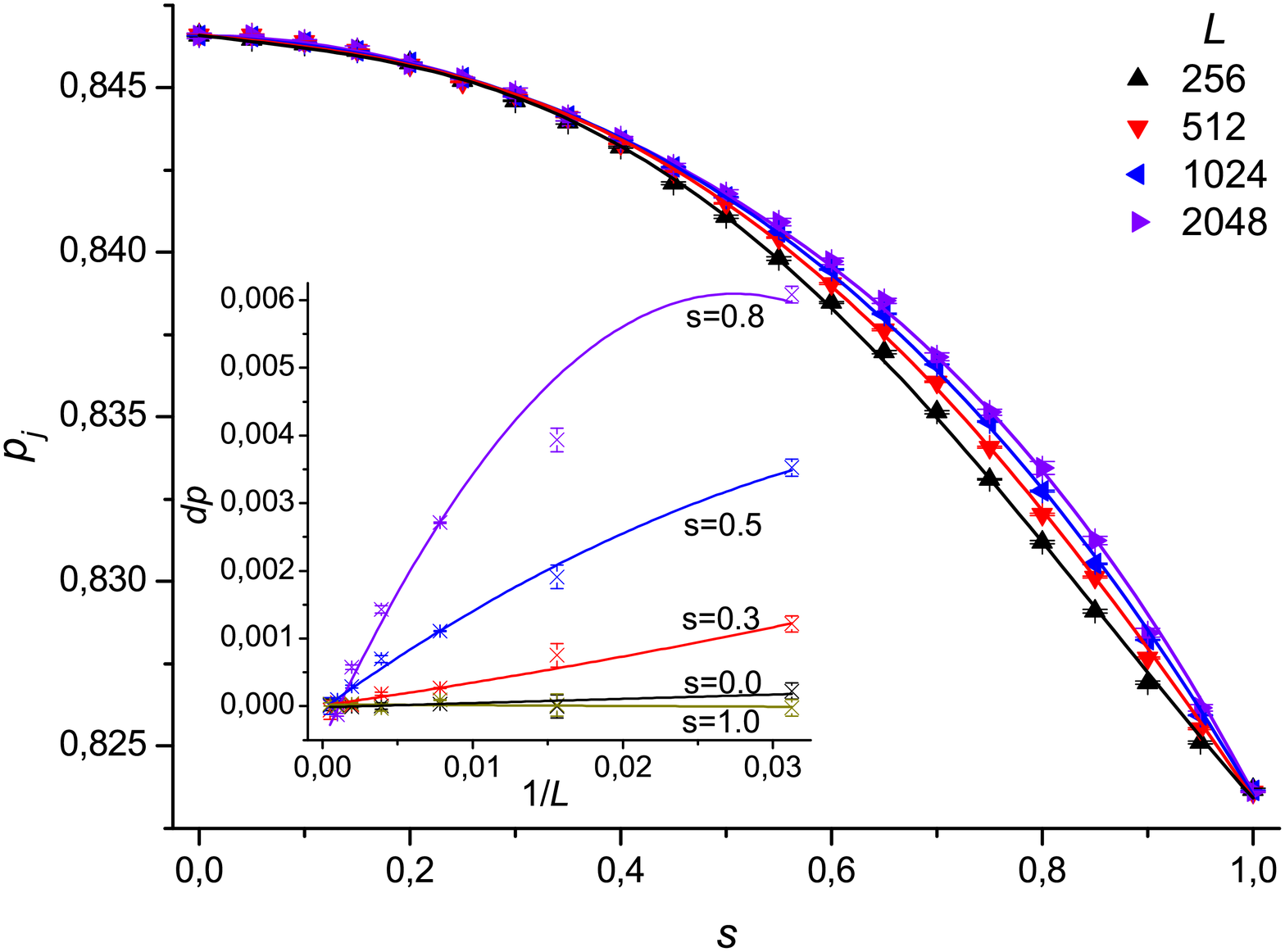}}\hfill
\subfigure[$k=8$ (RSA model)]{\includegraphics[width=0.5\linewidth]{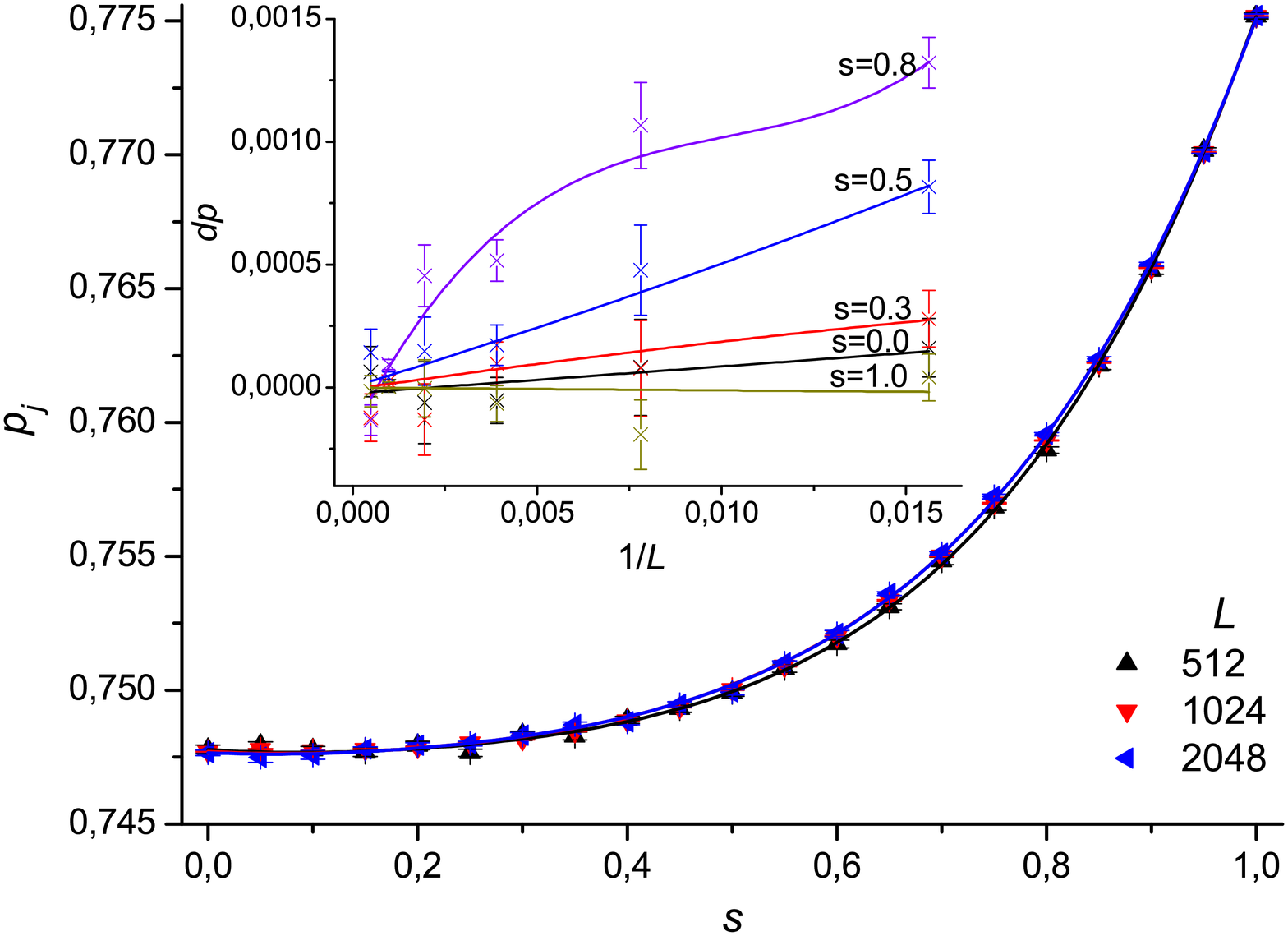}}\\
\subfigure[$k=3$ (RRSA model)]{\includegraphics[width=0.5\linewidth]{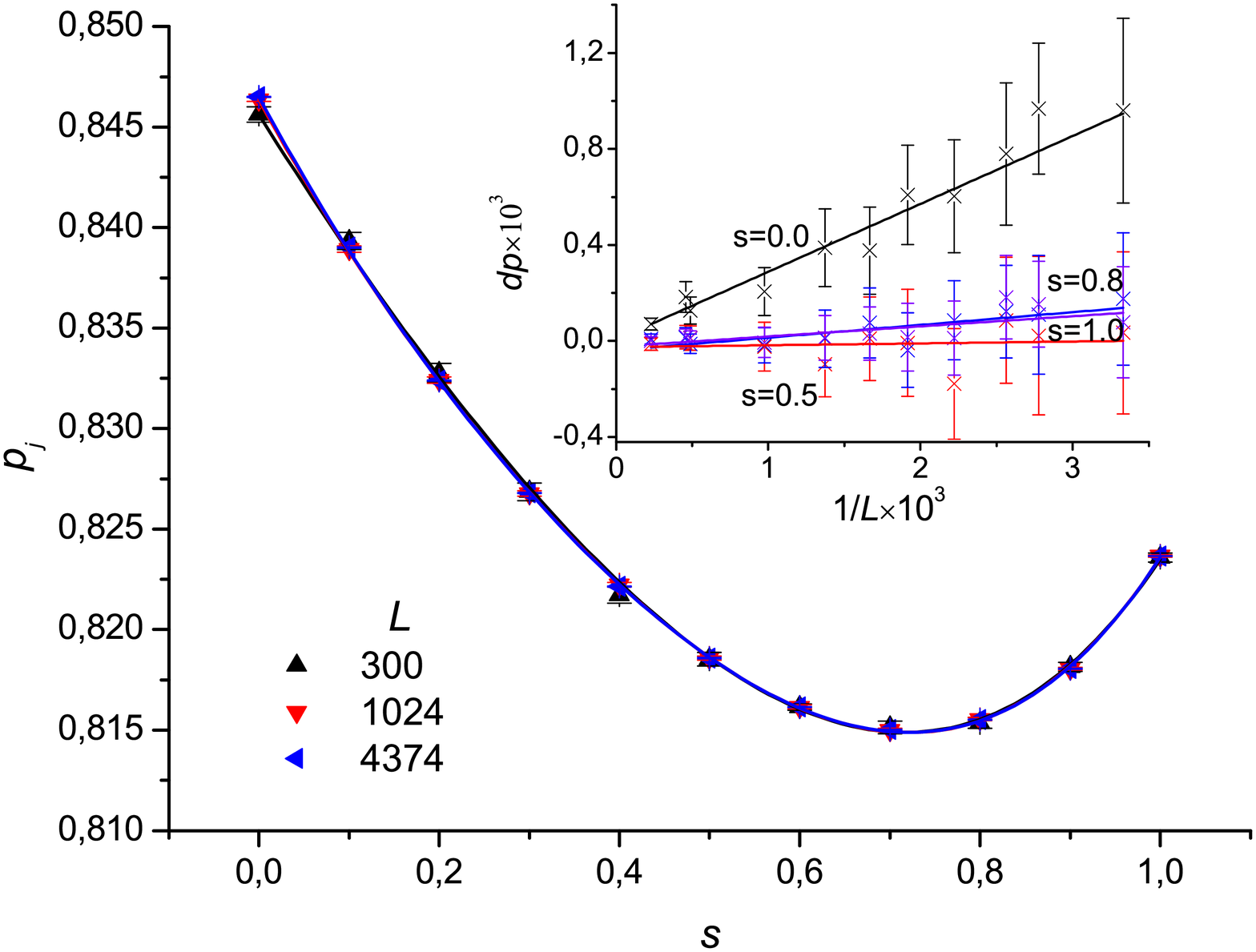}}\hfill
\subfigure[$k=8$ (RRSA model)]{\includegraphics[width=0.5\linewidth]{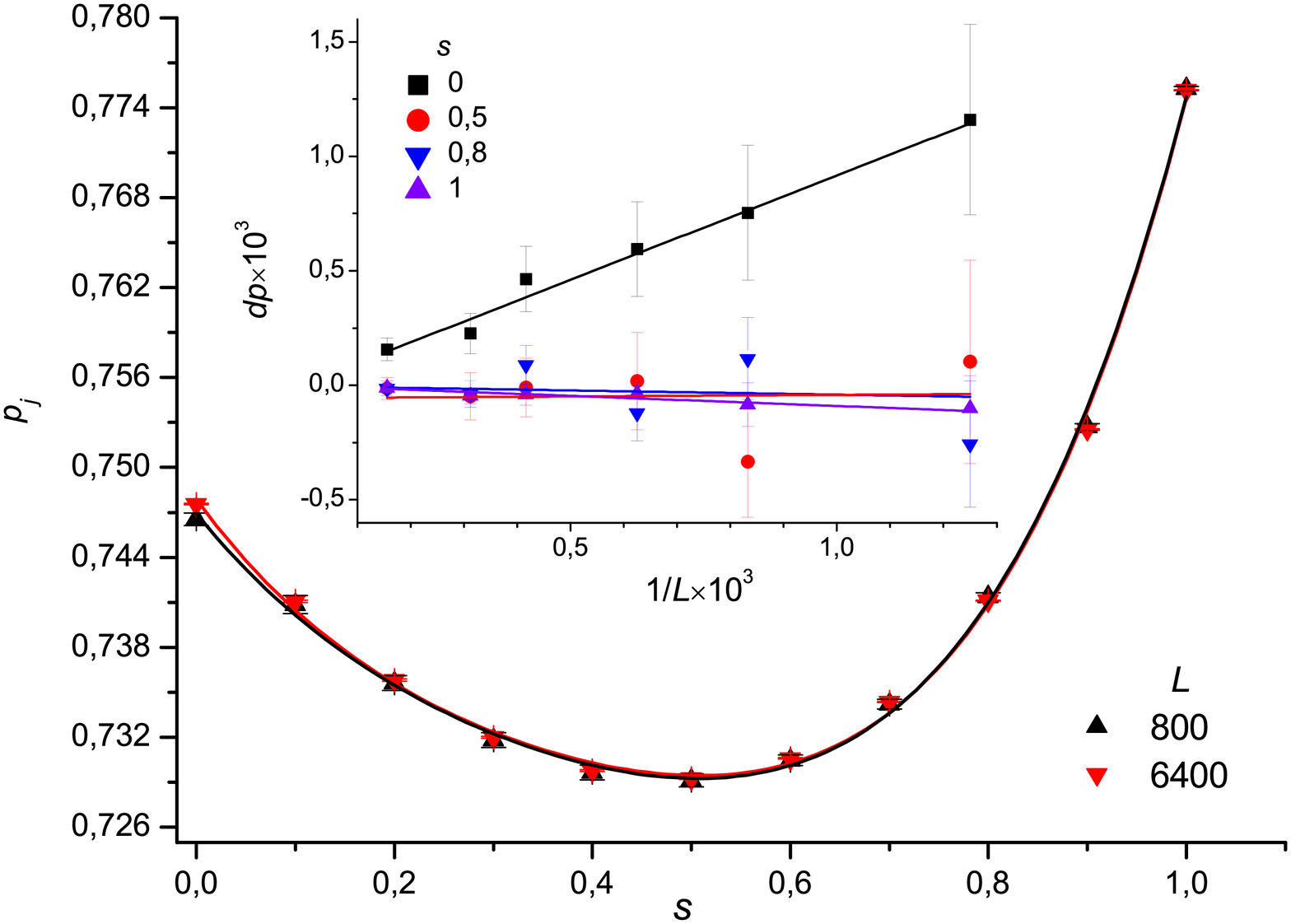}}
\caption{\label{fig:scalingRRSAk38}Jamming concentration $p$ \emph{vs.} order parameter $s$
at different values of $L$. Insets show
$dp=p_j(L\to\infty)-p_j$  \emph{vs.} $1/L$ for different values of
$s$. }
\end{figure*}

In our study, the length of $k$-mers varies between 2 and 128. To examine the finite size effect for RSA and RRSA models, different lattice sizes up to $L = 100k$ have been used.
The number of runs varies from 10 to 1000 depending of the lattice size $L$.

Two different random number generators have been applied for filling in the lattice with objects ($k$-mers) at given concentration and orientation. They are Mersenne Twister random number generator~\cite{Matsumoto} with a period of $2^{19937} - 1$ and the generator of Marsaglia \emph{et al.}~\cite{Marsaglia1990}. The results obtained using different generators are almost undistinguishable.

The connectivity of $k$-mers oriented in the vertical direction has been analyzed for jamming configurations, and the percolation threshold has been determined using the Hoshen-Kopelman algorithm~\cite{Hoshen1976}.
\begin{figure*}
\centering
\subfigure[RSA model]{\includegraphics[width=0.5\linewidth]{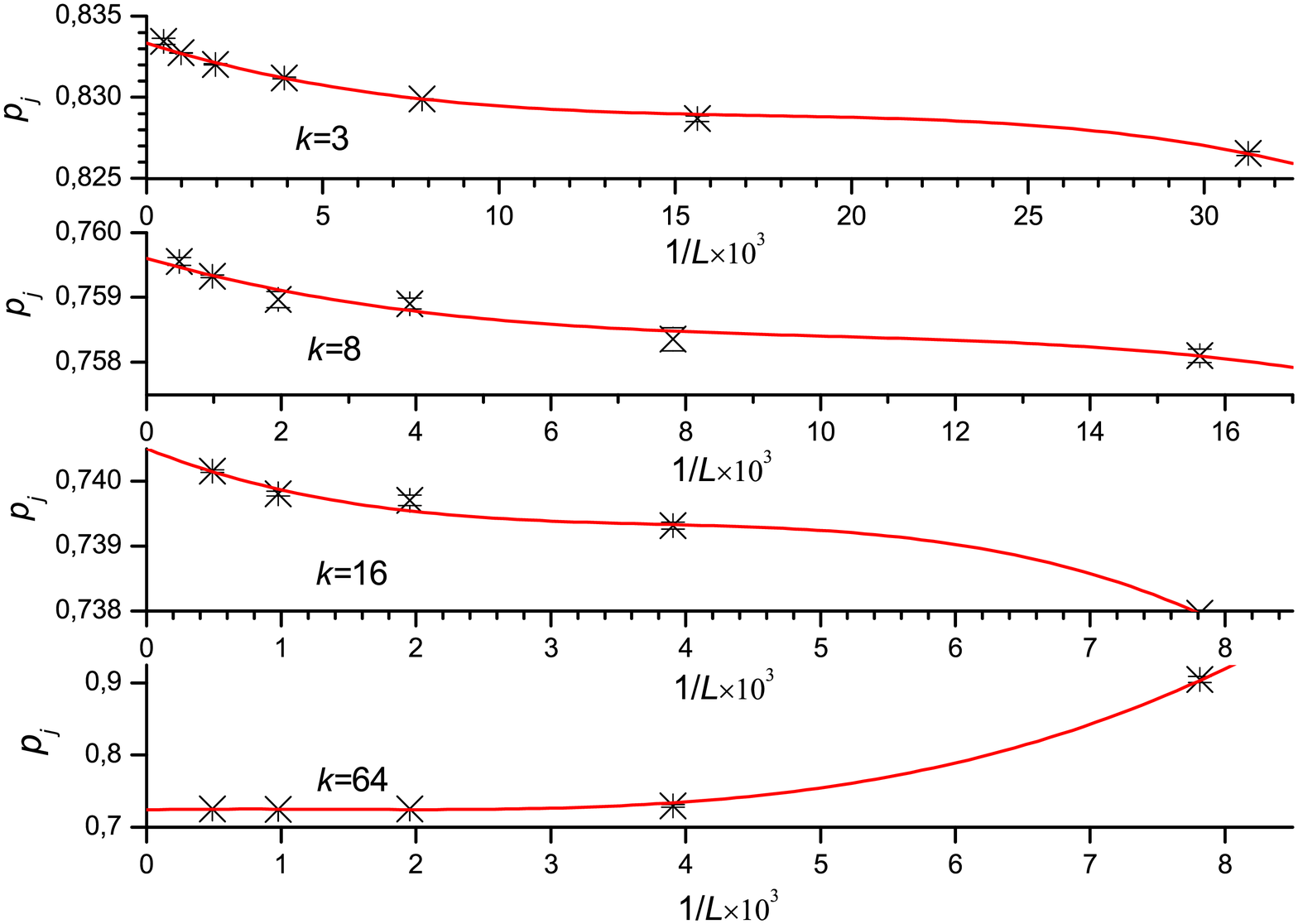}}\hfill
\subfigure[RRSA model]{\includegraphics[width=0.5\linewidth]{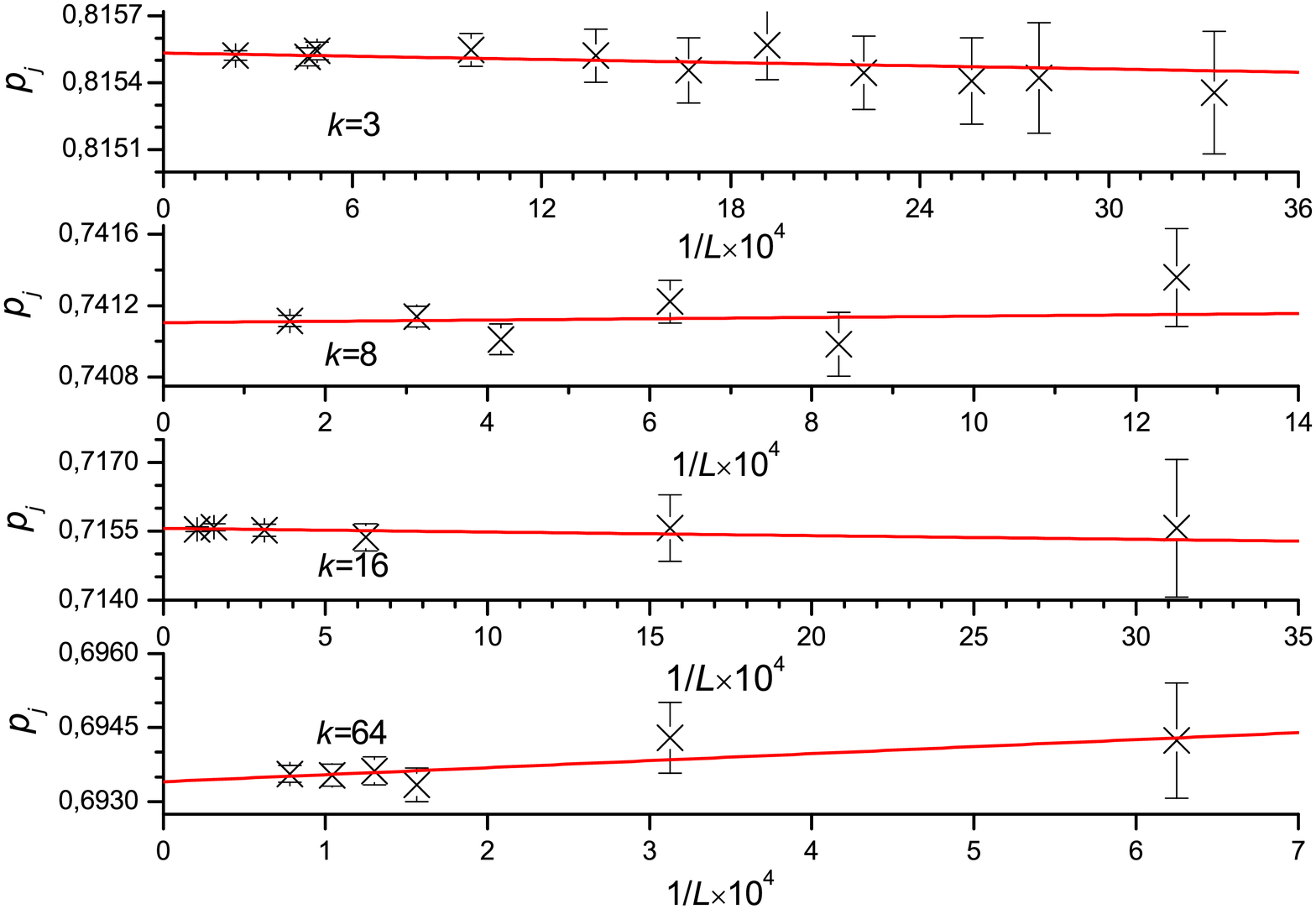}}
\caption {Jamming concentration $p_j$ \emph{vs.} $1/L$ for different
values of $k=3$(a), 8(b), 64(c) and fixed order parameter, $s=0.8$.} \label{fig:JCforfixeds}
\end{figure*}

\section{\label{sec:results}Results and discussion}

\subsection{\label{subsec:JP}Jamming configurations and its connectivity}

Jamming configurations obtained in the simulations are strongly dependent upon the order parameter and length of $k$-mers. However, visually, the difference in the structure of jamming patterns is not noticeable for the studied RSA and RRSA models. Examples of the typical jamming configurations for RRSA model at different values of order parameter $s$ are presented in Fig.~\ref{fig:djm_m2}.

For randomly oriented linear $k$-mers (\emph{i.e.}, at $s=0$) the typical domain structure in form of blocks of parallelly oriented $k$-mers has been observed. These blocks can be represented as the squares of size $k \times k$~\cite{Vandewalle}. One can present a jamming configuration as a combination of:
\begin{itemize}
  \item blocks of vertically oriented $k$-mers ($v$-blocks);
  \item blocks of horizontally oriented $k$-mers ($h$-blocks);
  \item empty sites (voids).
\end{itemize}

The relative area occupied by $v$- and $h$-blocks is approximately the same at $s=0$ (Fig.~\ref{fig:djm_m2}(a)).
Increase of $s$ results in decrease of the relative area occupied by $h$-blocks and $v$-blocks become dominating in jamming patterns. Moreover, the relative area occupied by $h$-blocks is visually larger for RRSA model than for RSA model (compare Fig.~\ref{fig:djm_m2}(b),(c)) with
Fig.~\ref{fig:djm_m2}(e),(f)).  Finally, at $s=1$, the jamming configuration transfers into the 1d-like domains (independent parallel 1d jamming lines of $k$-mers)(Fig.~\ref{fig:djm_m2}(d)).

It is interesting that the infinite connectivity between the similar $v$- or $h$-blocks (\emph{i.e.},  percolation) is not observed for randomly oriented linear $k$-mers (at $s=0$). The visual observations of jamming patterns show that connectivity between $v$-blocks increases and between $h$-blocks decreases with increase of $s$. It can be easily demonstrated by analyzing the structure of the largest cluster of vertically oriented $k$-mers (Fig.~\ref{fig:djm_p2}). At small $s$ (\emph{e.g.}, at $s=0.05$ in Fig.~\ref{fig:djm_p2}(a), (d)), the connectivity of $v$-blocks is limited and the largest cluster occupies only the finite part of the lattice. However, the size of the largest cluster is higher for RSA model than for RRSA model. Increase of $s$ results in growth of the size of the largest cluster, and this cluster spans through the lattice at some threshold value of $s_c$ (see, \emph{e.g.}, Fig.~\ref{fig:djm_p2}(b) for RSA model and Fig.~\ref{fig:djm_p2}(f) for RRSA model). Examples of the percolation probability $f$ \emph{vs.} the order parameter $s$ at $k=2,8$ and $L=1024$ are presented in Fig.~\ref{fig:f vs s}. The threshold value of $s_c$ that corresponds to the percolation of vertically oriented $k$-mers is determined from the condition of $f=0.5$.

The usual finite size scaling analysis of the percolation behavior is done, and it is obtained that $s(L)$ follows scaling law governed by the universal scaling exponent~$\nu$:
\begin{equation}\label{Eq:e2p}
    \left| {{s_{c}}(L) - {s_{c}}(\infty )} \right|\propto {L^{ - \frac{1}{\nu }}},
\end{equation}
where $\nu = 4/3$ is the critical exponent of correlation length for 2d random percolation problem~\cite{Stauffer}.

For the thermodynamic limit ($L\to\infty$), dependencies of $s_c$ () \emph{vs.} length of $k$-mer for RSA and RRSA models are presented in Fig.~\ref{fig:sc_vs k}; the inset of this figure shows examples of finite size scaling dependencies for $k=2$ and $k=8$. The observed percolation behavior of the vertically oriented $k$-mers is rather different for RSA and RRSA model, which evidently reflects the difference in the structure of jamming patterns of these models. For the same $k$, the RSA model gives lower value $s_c$ than RRSA model, \emph{e.g.}  $k=16$,  $s_c\approx 0.126$ (RSA model) and $s_c\approx 0.240$ (RRSA model). Note that in both models for the dimers ($k=2$), the threshold values of $s_c$ are rather close to $\approx 0.21$. However, the value of $s_c$ decreases as the length of $k$-mer increases for RSA model, and opposite behavior is observed for RRSA model. Finally, in continuous limit $k\to\infty$, the difference between the threshold order parameters of RSA and RRSA models becomes rather large, $\triangle s_c\approx 0.12$. The difference in behavior of $s_c(k)$ observed for RSA and RRSA models evidences higher connectivity between $v$-blocks for RSA model than for RRSA model, and this tendency increases as the length of $k$-mer increases.

\begin{table*}[!tb] 
\caption{\label{tab:pjvsSk}
Estimated jamming concentrations, $p_j=p_j(L \to \infty)$, as a function of order parameter, $s$, at different values of $k$ for RSA and RRSA models.
Data 1d correspond to exact results that are obtained by direct calculation using Eq.\ref{eq:ue1Djamm}.}
\begin{tabular}{|c||c|c||c|c||c|c||c|c||c|c|}
  \hline
&\multicolumn{2}{|c||}{$k$=2} & \multicolumn{2}{|c||}{$k$=3} & \multicolumn{2}{|c||}{$k$=4} & \multicolumn{2}{|c||}{$k$=8}  \\ \hline
$s$  &RSA    & RRSA  & RSA   & RRSA  & RSA   & RRSA  & RSA   & RRSA  \\\hline
0.0& 0.906(8)& 0.905(9)& 0.846(6)& 0.845(5)& 0.810(4)& 0.809(1)& 0.747(6)& 0.746(8)\\
0.1& 0.906(7)& 0.899(4)& 0.846(4)& 0.838(8)& 0.810(2)& 0.802(8)& 0.747(6)& 0.740(8)\\
0.2& 0.906(3)& 0.892(9)& 0.845(8)& 0.832(7)& 0.809(5)& 0.796(9)& 0.747(9)& 0.736(0)\\
0.3& 0.905(8)& 0.887(1)& 0.844(8)& 0.822(7)& 0.808(7)& 0.792(2)& 0.748(0)& 0.732(3)\\
0.4& 0.905(3)& 0.882(2)& 0.843(5)& 0.821(7)& 0.807(6)& 0.787(6)& 0.748(8)& 0.729(5)\\
0.5& 0.904(2)& 0.877(1)& 0.841(9)& 0.818(7)& 0.806(3)& 0.785(3)& 0.750(0)& 0.729(1)\\
0.6& 0.902(6)& 0.873(0)& 0.839(9)& 0.816(4)& 0.805(0)& 0.784(2)& 0.752(3)& 0.730(8) \\
0.7& 0.898(5)& 0.869(7)& 0.837(1)& 0.815(0)& 0.803(6)& 0.784(6)& 0.755(2)& 0.734(3)\\
0.8& 0.892(5)& 0.867(2)& 0.833(8)& 0.815(5)& 0.802(3)& 0.787(3)& 0.759(7)& 0.741(0)\\
0.9& 0.882(3)& 0.865(4)& 0.828(7)& 0.818(0)& 0.801(7)& 0.792(9)& 0.766(0)& 0.752(5)\\
\hline
1.0&\multicolumn{2}{|c||}{0.86466(7)} & \multicolumn{2}{|c||}{0.82365(3)} & \multicolumn{2}{|c||}{0.80389(6)} & \multicolumn{2}{|c||}{0.77518(5)}  \\ \hline
1d&\multicolumn{2}{|c||}{0.864665717} & \multicolumn{2}{|c||}{0.823652963} & \multicolumn{2}{|c||}{0.803893480} & \multicolumn{2}{|c||}{0.775184833}  \\
\hline \hline
&\multicolumn{2}{|c||}{$k$=16} & \multicolumn{2}{|c||}{$k$=32} & \multicolumn{2}{|c||}{$k$=64} & \multicolumn{2}{|c||}{$k$=128}
\\ \hline
$s$  &RSA    & RRSA  & RSA   & RRSA  & RSA   & RRSA  & RSA   & RRSA
\\\hline
0  & 0.710(9)& 0.709(3)& 0.689(4)& 0.687(9)& 0.678(2)& 0.674(3)& 0.668(9)& 0.666(3)\\
0.1& 0.711(2)& 0.704(3)& 0.690(4)& 0.683(8)& 0.680(0)& 0.670(9)& 0.668(0)& 0.663(2)\\
0.2& 0.712(4)& 0.700(2)& 0.692(4)& 0.680(5)& 0.682(4)& 0.667(6)& 0.672(8)& 0.660(2)\\
0.3& 0.714(6)& 0.697(4)& 0.695(6)& 0.677(4)& 0.686(0)& 0.665(4)& 0.677(2)& 0.659(4) \\
0.4& 0.717(1)& 0.696(2)& 0.699(9)& 0.677(6)& 0.691(2)& 0.666(1)& 0.684(2)& 0.659(9)\\
0.5& 0.720(8)& 0.697(0)& 0.705(6)& 0.678(6)& 0.697(6)& 0.668(6)& 0.691(6)& 0.662(3)\\
0.6& 0.726(0)& 0.700(3)& 0.712(4)& 0.683(2)& 0.705(6)& 0.673(3)& 0.700(2)& 0.667(5)\\
0.7& 0.732(3)& 0.705(7)& 0.720(7)& 0.690(2)& 0.713(8)& 0.681(4)& 0.711(0)& 0.676(6)\\
0.8& 0.740(3)& 0.715(6)& 0.730(6)& 0.701(5)& 0.724(8)& 0.693(5)& 0.723(0)& 0.688(8)\\
0.9& 0.749(8)& 0.730(7)& 0.741(9)& 0.719(1)& 0.737(4)& 0.712(8)& 0.736(5)& 0.709(4)\\
\hline
1.0&\multicolumn{2}{|c||}{0.7612(4)} & \multicolumn{2}{|c||}{0.7543(9)} & \multicolumn{2}{|c||}{0.7509(7)} & \multicolumn{2}{|c||}{0.7493(0) }  \\ \hline
1d&\multicolumn{2}{|c||}{0.761250552} & \multicolumn{2}{|c||}{0.754388934} & \multicolumn{2}{|c||}{0.750984590} & \multicolumn{2}{|c||}{0.749289044}  \\
\hline
\end{tabular}
\end{table*}

\begin{figure}
\centering
\includegraphics[width=\linewidth]{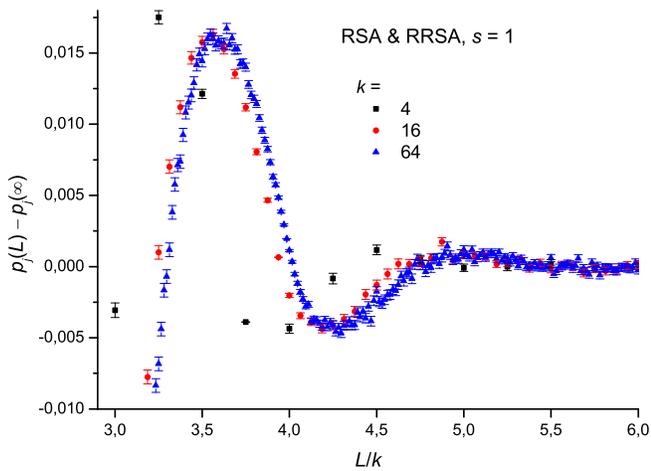}
\caption{Difference of jamming concentrations $p_j(L)-p_j(\infty)$ \emph{vs.} $L/k$ ratio
for completely ordered $k$-mers, ($s=1$) at different values of $k$(RSA and RRSA
models).\label{fig:1dscal}}
\end{figure}

\begin{figure*}
\centering
\subfigure[RSA model]{\includegraphics[width=0.5\linewidth]{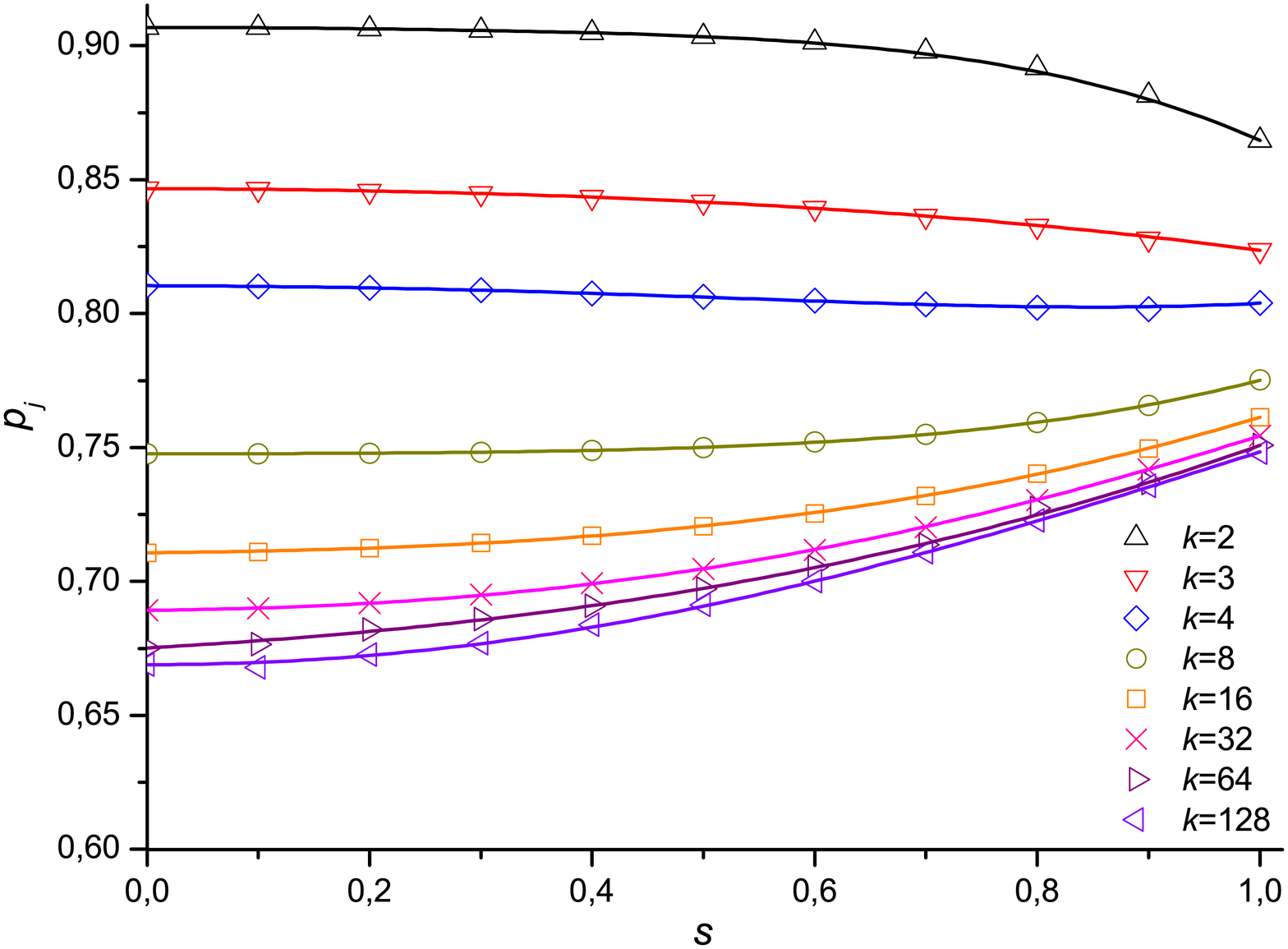}}\hfill
\subfigure[RRSA model]{\includegraphics[width=0.5\linewidth]{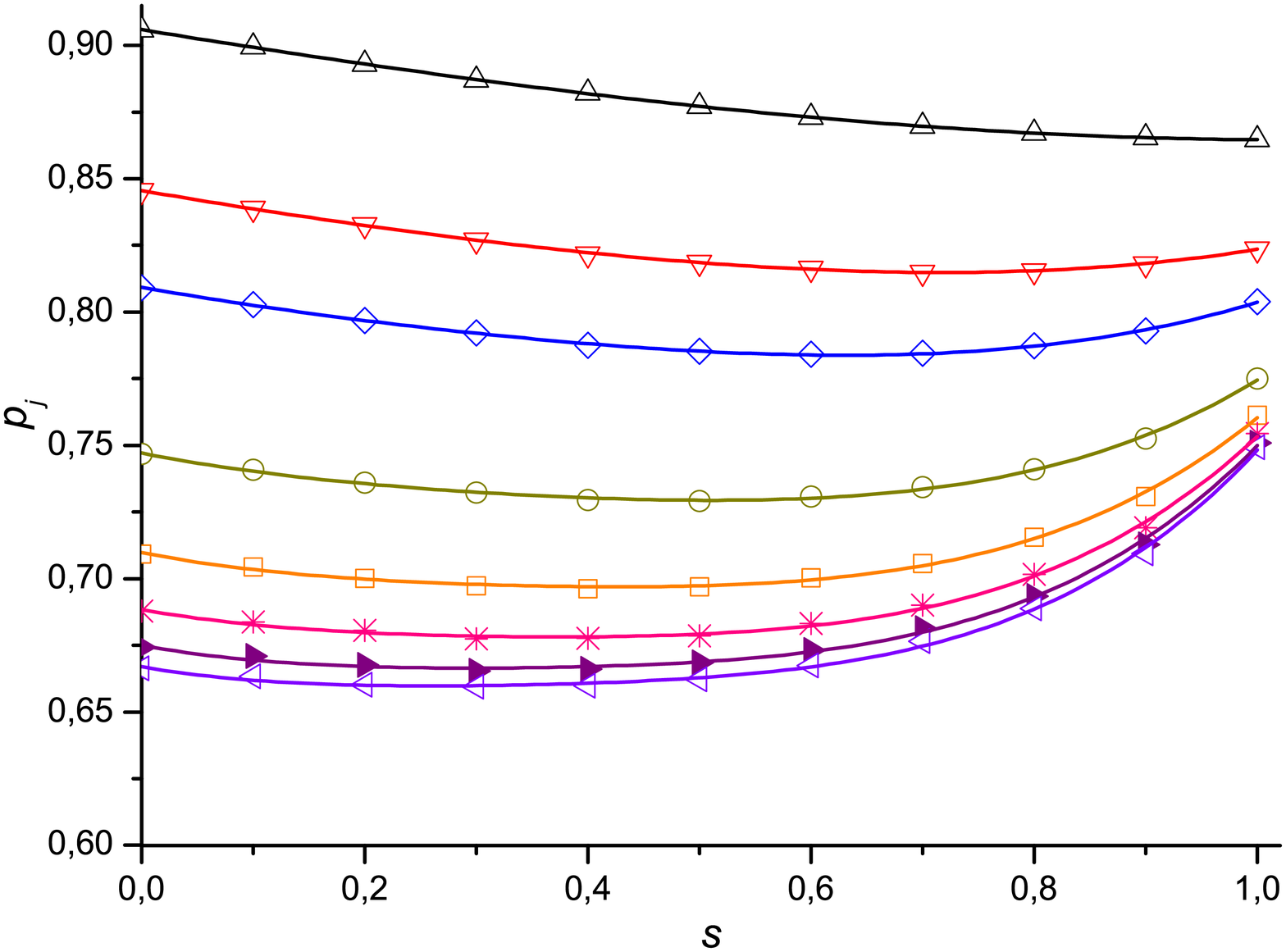}}
\caption {Jamming concentration $p_j=p_j(L \to \infty)$ \emph{vs.} order parameter $s$ for different values of $k$.} \label{fig:PvsS}
\end{figure*}

\begin{figure*}
\centering
\subfigure[]{\includegraphics[width=0.5\linewidth]{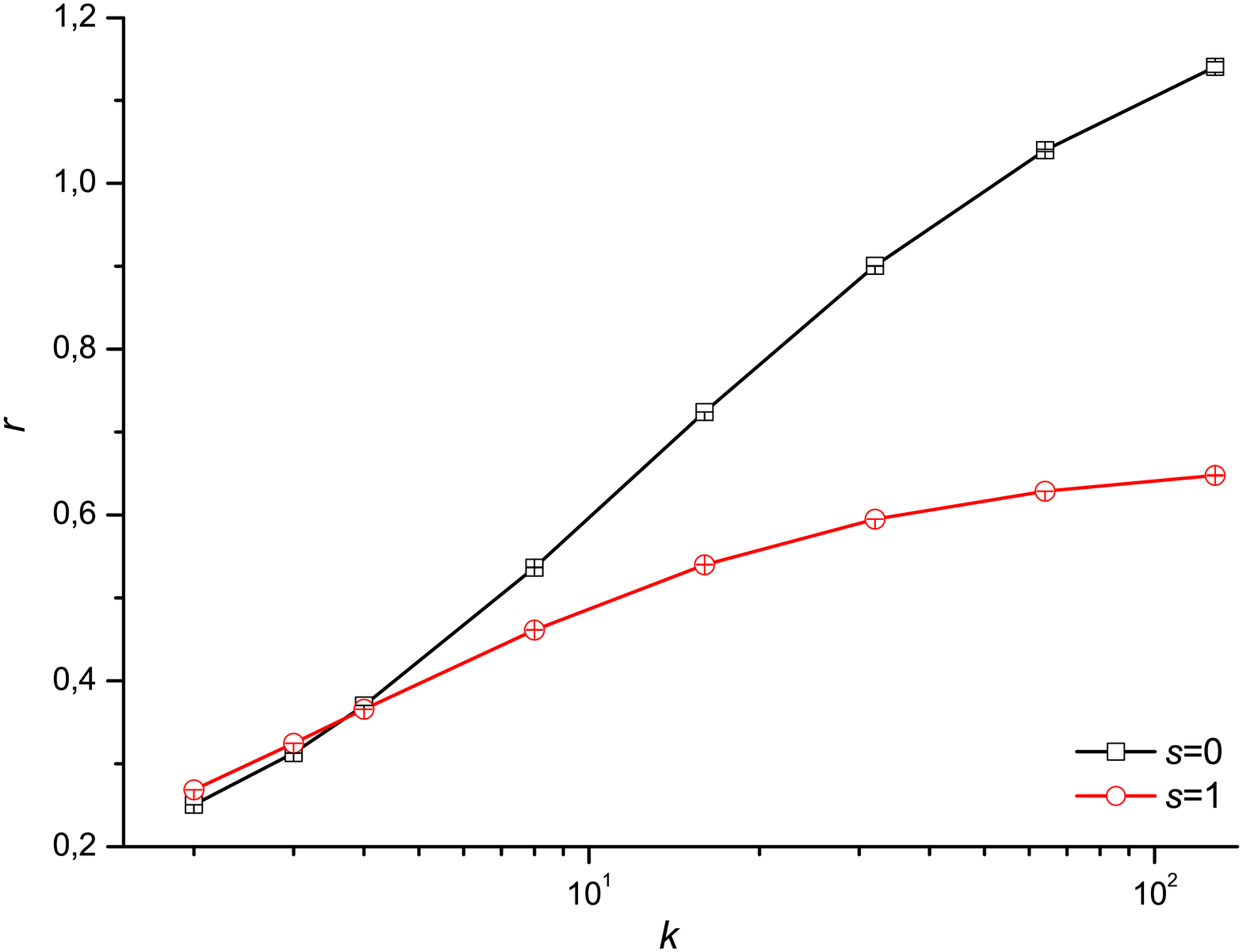}}\hfill
\subfigure[]{\includegraphics[width=0.5\linewidth]{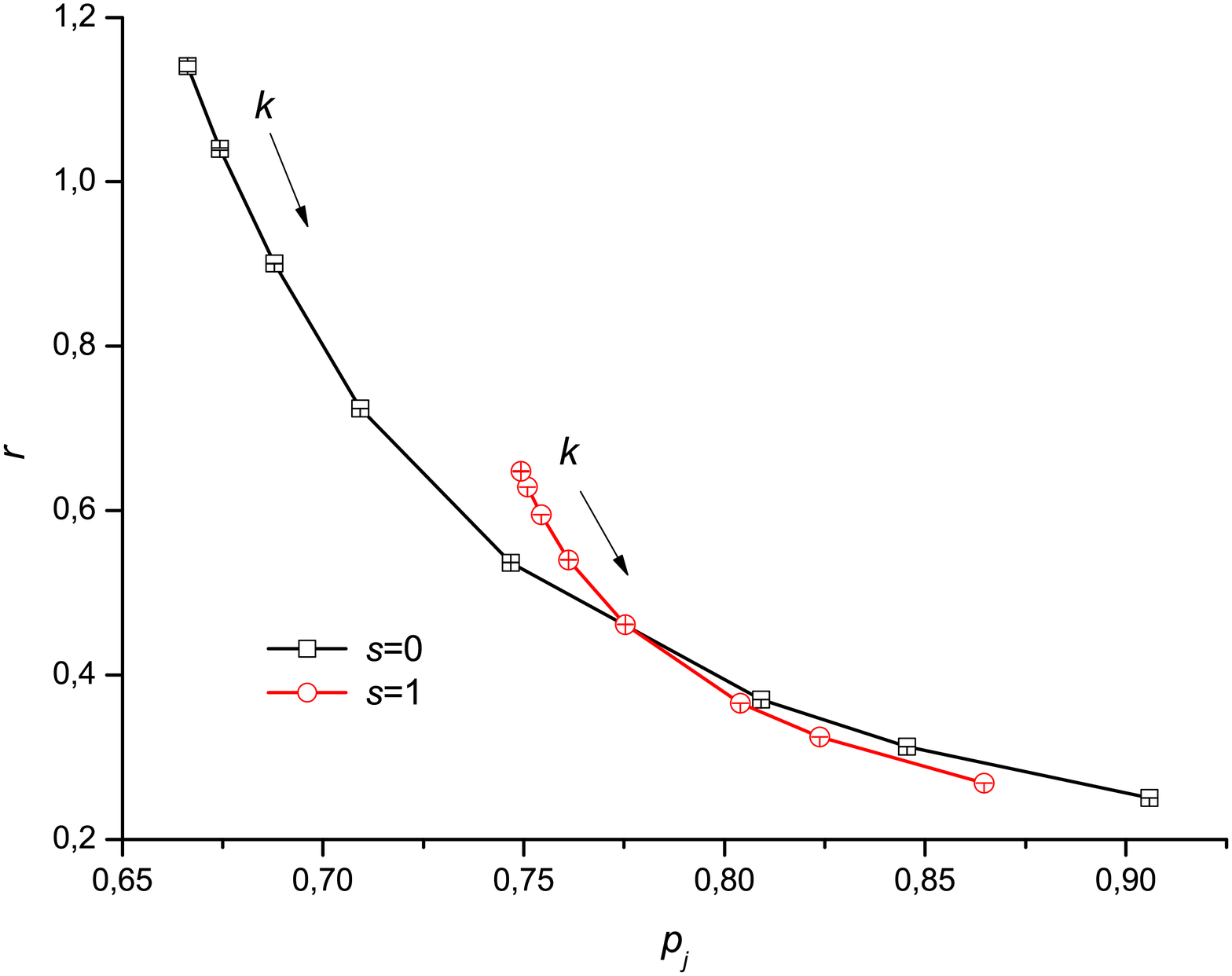}}
\caption {Effective pore radius $r$ \emph{vs.} value of $k$ (a) and jamming concentration $p_j=p_j(L \to \infty)$ (b) at $s=0$ and $s=1$ (RSA and RRSA models).} \label{fig:porek}
\end{figure*}

\begin{figure}[h]
\centering
\includegraphics[width=\linewidth]{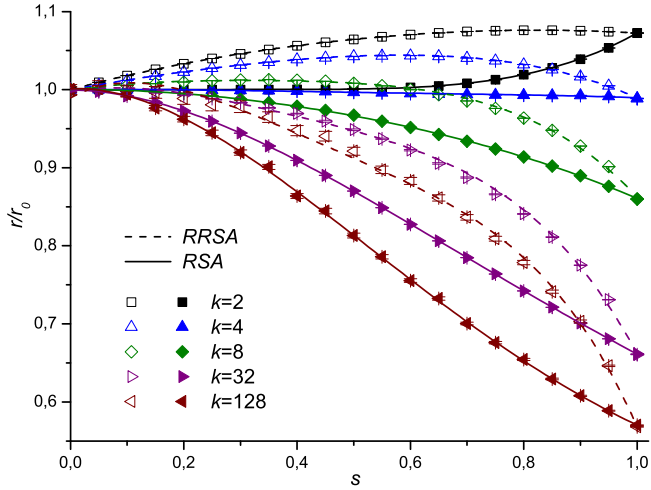}
\caption {Relative effective pore radius $r/r_0$ \emph{vs.} order parameter $s$ at different values of $k$ for RSA and RRSA models). Here, $r_o=r(s=0)$.} \label{fig:pores}
\end{figure}

\begin{figure} 
\centering
\includegraphics[width=\linewidth]{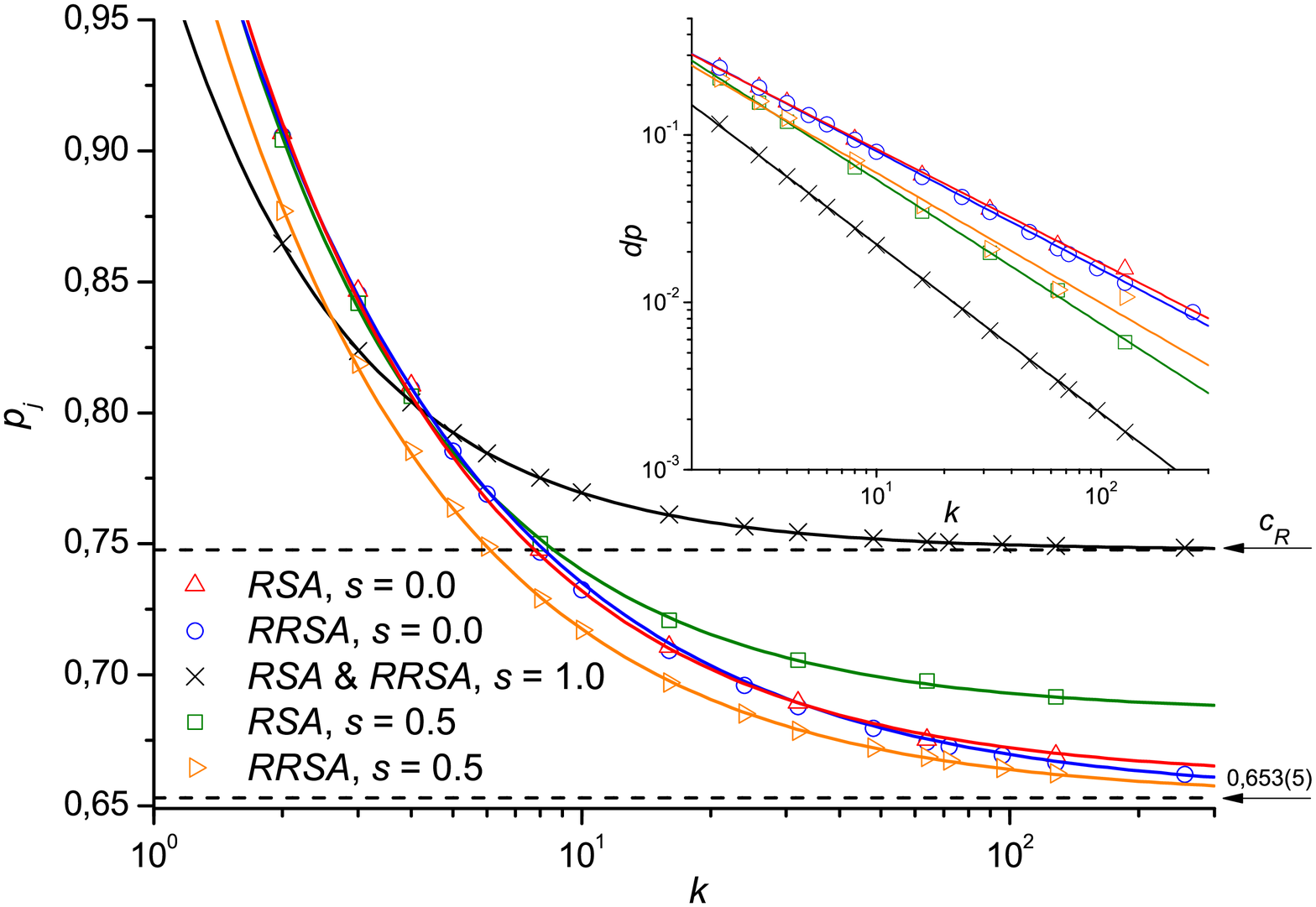}
\caption{Jamming concentration $p_j$ \emph{vs.} $k$ at
different values of order parameter $s$. Inset shows
$dp=p_j-p_j(k\to\infty)$ \emph{vs.} $k$, the different slopes
correspond to the different scaling exponents $\alpha$. (RSA and RRSA
models).\label{fig:Pvsk_pow}}
\end{figure}

\subsection{\label{subsec:FSE}Finite-size effects}

Commonly, it is assumed that finite-size effects on the jamming coverage of a periodic lattice decrease very rapidly with the lattice size $L$ increase \cite{BarteltJMP1991, BrosilowPRA1991}. For isotropic problem of adsorption of $k$-mers on a square lattice ($s=0$), it was observed that standard deviation of the coverage is a linear function of $k/L$ and it become negligible in the limit of $k/L\to 0$  \cite{Bonnier1994PRE}.

Our investigations show that in case of anisotropic problem the finite-size effects, as well as $k$ and $s$, are rather sensitive to the type of adsorption model (RSA or RRSA). Fig.~\ref{fig:scalingRRSAk38} demonstrates examples of jamming concentration $p_j$ \emph{vs.} $s$ for both models and different lattice sizes $L=32$--$2048$ in two particular cases when $k=3$ and $k=8$. For clearer demonstration of the finite-size effects on the jamming concentration, they are also represented in the form of $dp=p_j(L\to\infty)-p_j(L)$ \emph{vs.} the inverse lattice size $1/L$ (see insets to Fig.~\ref{fig:scalingRRSAk38}).

In RSA model, (Fig.~\ref{fig:scalingRRSAk38}(a),(b) the finite size effects are rather small in the limits of $s\to 0$ and $s\to 1$. However, for the intermediate anisotropy $s$, they are large and $dp=p_j(L\to\infty)-p_j$ \emph{vs.} $1/L$ looks like nonlinear even for the large lattice sizes $L$. From the other side, in RRSA model (Fig.~\ref{fig:scalingRRSAk38}(c),(d)) the finite size effects are very small for any $k$ and $s$ with the exception of the limit of isotropic problem, $s\to 0$.

The differences between the finite size effects of RSA and RRSA models at intermediate anisotropy ($s=0.8$) and different values of $k$ are presented in Fig.~\ref{fig:JCforfixeds} in the form of $p_j$ \emph{vs.} $1/L$ dependencies. For RSA model, these dependencies look like nonlinear even for large lattice sizes, and this fact obstructs application of the scaling relation in the thermodynamic limit ($L\to\infty$)~\cite{Stauffer}
\begin{equation}\label{Eq:e2}
    \left| {{p_{j}}(L) - {p_{j}}(\infty )} \right|\propto {L^{ - \frac{1}{\nu }}},
\end{equation}
where $\nu = 1.0 \pm 0.1$~\cite{Vandewalle}.

\begin{table*}[!tb] 
\caption{\label{tab:pjvsSkoo}
Estimated parameters $p_{j}(\infty)$, $a$ and, $\alpha$ in a power law function Eq.~\ref{eq:power} using for
the fitting of a dependence of jamming concentration $p_j$ \textit{vs.} length of linear segment $k$. The data are presented for RSA and RRSA models at different values of $s$. Data 1d, exact, corresponds to fitting exact values that are obtained by direct calculation using Eq.\ref{eq:ue1Djamm}.
In all cases the coefficient of determination $\rho$ was higher than $0.9997$.}
\begin{tabular}{|c||c|c||c|c||c|c||c|c|
}
  \hline
&\multicolumn{2}{|c||}{$p_{j}(k=\infty)$} & \multicolumn{2}{|c||}{$a$} & \multicolumn{2}{|c||}{$\alpha$}
\\ \hline
$s$  &RSA    & RRSA  & RSA   & RRSA  & RSA   & RRSA
\\\hline
0  &0.655(9)&0.652(8)&0.416(0)&0.417(5)&0.720(7)&0.713(7)\\
0.1&0.656(5)&0.650(8)&0.414(5)&0.414(9)&0.720(7)&0.730(1)\\
0.2&0.661(9)&0.648(5)&0.412(6)&0.411(1)&0.747(2)&0.741(3)\\
0.3&0.667(9)&0.648(0)&0.408(9)&0.405(4)&0.774(1)&0.753(2)\\
0.4&0.676(5)&0.649(6)&0.406(2)&0.397(0)&0.820(7)&0.766(6)\\
0.5&0.685(8)&0.652(9)&0.401(3)&0.385(7)&0.870(9)&0.776(1)\\
0.6&0.696(3)&0.659(1)&0.393(8)&0.370(9)&0.928(6)&0.788(2)\\
0.7&0.707(8)&0.669(1)&0.380(1)&0.352(8)&0.991(9)&0.809(8)\\
0.8&0.721(0)&0.683(0)&0.359(1)&0.328(2)&1.063(9)&0.830(8)\\
0.9&0.735(1)&0.705(0)&0.321(7)&0.295(0)&1.128(3)&0.876(5)\\
\hline
1.0&\multicolumn{2}{|c||}{0.747(2)} & \multicolumn{2}{|c||}{0.235(3)} & \multicolumn{2}{|c||}{1.016(6)}
\\ \hline
1d, exact&\multicolumn{2}{|c||}{0.74759792} & \multicolumn{2}{|c||}{0.227(7)} & \multicolumn{2}{|c||}{1.011(1)}
\\
\hline
\end{tabular}
\end{table*}

As a result, we have to utilize rather large lattices ($L>512$--$2048$) and nonlinear relations for $p(1/L)$ in a form of polynomials of degree 3 for extrapolation of the results to thermodynamic limit ($L\to\infty$) (see, \emph{e.g.}, Fig.~\ref{fig:JCforfixeds}). For RSA model, this technique allows to get rather reliable estimations of jamming concentrations at $L\to\infty$ with the error bar not exceeding $\pm0.002$.

For RRSA model, the finite size effects are very small at any $L/k \gg 1$ and $s$ (with the exception of the limit of isotropic problem, $s\to 0$); therefore, the final investigations were performed for different $k$ with $L = 100k$ and $100$ independent runs. Our estimation of the error bar for $p(L\to\infty)$ is about $ \pm 0.0001$. For the particular case of the limit of isotropic problem, $s\to 0$ the procedure used for estimation of $p(L \to \infty)$ is the same as for RSA model.

Note that at moderate ratio $L/k<5$ the finite size effects may be rather complex. The examples of $p_j(L)-p_j(\infty)$ \emph{vs.} $L/k$ dependencies for completely ordered $k$-mers, ($s=1$) at different values of $k$  are presented in Fig.~\ref{fig:1dscal}). The observed noticeable oscillating scaling behavior evidently reflects commensurability of $k$ and $L$ values.

The precision of the estimation of $p(L \to \infty)$ is tested for the limit cases of completely anisotropic ($s=1$) and isotropic ($s=0$)  problems.

The particular case of complete alignment ($s=1$) corresponds to the simplified 1d problem when RSA and RRSA models are indistinguishable.
The data presented in Table~\ref{tab:pjvsSk} show also that our results for both RSA and RRSA models at $s=1$ are very close to the analytical results calculated 1d problem from Eq.~\eqref{eq:ue1Djamm}. The numerically obtained data are precise within 4--5 significant digits.

For isotropic problem ($s=0$), both RSA and RRSA models give very close estimations of $p_j(k)$ value. In this case, our results for $p_j$ for small $k$-mers ($k = 2$--$8$) (see Table~\ref{tab:pjvsSk}) are in a reasonable correspondence with the previously published data, \textit{e.g.},
0.9068~\cite{NordJPC1985}, 0.9067(7)~\cite{deOliveiraPRA1992}, 0.906~\cite{Vandewalle} ($k=2$), 0.8465~\cite{Evans}, 0.847~\cite{Vandewalle} ($k=3$), 0.811~\cite{Vandewalle} ($k=4$) and  0.757~\cite{Vandewalle} ($k=8$).

The similar jamming behavior of the RSA and RRSA models in the isotropic case $s=0$ is expected. For RSA model in the isotropic case, rejection of the unsuccessful attempt is followed by the next choice of $k$-mers with random orientation. In this case, the RSA model also allows preservation of the predetermined order parameter. That is a reason why the RSA and RRSA models give the similar estimations for the values of $p_j(L \to \infty)$. However, the amplitude of order parameter fluctuations during the deposition of $k$-mers may be rather different for RSA and RRSA models, and in fact it results in a noticeable difference of the finite-size scaling effects at $s=0$ observed for RSA and RRSA models (see Fig.~\ref{fig:scalingRRSAk38}).

Finally, the data on $p(L \to \infty)$ \emph{vs.} $s$ dependencies for RSA and RSSA models at different $k$ obtained in a result of the described scaling analysis are presented  in Table~\ref{tab:pjvsSk}.

\subsection{Jamming concentration}

\subsubsection{Jamming concentration vs. order parameter}
Fig.~\ref{fig:PvsS} demonstrates behavior of jamming concentrations, $p_j=p_j(L \to \infty)$ , as a function of order parameter, $s$, for RSA and RRSA models, respectively.

The $p_j(s)$ dependencies are rather different for short and long $k$-mers. For instance, if $k$-mers are short ($k\leqslant 4$), more dense configurations are observed for disordered systems ($s=0$) as compared to those of completely ordered case ($s=1$). However, an opposite behavior is observed for longer $k$-mers.
For RSA model, the value of $p_j(k)$ monotonically decreases ($k\leqslant 4$) or increases ($k>4$) when the order parameter $s$ increases.
For RRSA model, the value of $p_j(k)$ goes through the minimum at $k\geqslant 3$ (RRSA-model) when $s$ increases.

At each value of $k$, the jamming concentrations $p_j$ are practically the same at $s=0$ and $s=1$ for the both RSA and RRSA models.
However, in the intermediate region of $s$ we observe $p_j(s)$(RSA)$>p_j(s)$(RRSA). The maximal difference between $p_j(s)$(RSA) and $p_j(s)$(RRSA) is observed at $s\approx 0.5$--$0.7$.

Such behavior may be explained as follows. In RSA model, the substrate "selects" the $k$-mer with appropriate orientation. It results in amplification of the actual order parameter $s_0>s$ (Fig.~\ref{fig:s_for_iter}).

Such amplification reflects intensive nucleation of $v$-domains at initial stages of deposition. At $s>0$, the $v$-domains merge into percolating clusters when $s>s_c$. For well oriented systems (at $s\approx 0.5-0.7$), the jamming structures in RSA model are large 1d-like $v$-domains with small inclusions of $h$-domains and voids.

In RRSA model, the order parameter keeps the level of approximately $s_0\approx
s$ (Fig.~\ref{fig:s_for_iter}). The packing of $k$-mers in the form of $h$-and $v$-domains arising at initial stages of RRSA deposition is sparser than for RSA model.  As a result, the $v$-domains merge into percolating clusters at higher level of $s_c$ in RRSA model than in RSA model. Finally, for the well oriented systems (at $s\approx 0.5$--$0.7$), the jamming structures of RRSA model include higher number of $h$-domains with large voids between them than those of RSA model.

The effective pore radius $r$ in the jammed structures is calculated as
\begin{equation}\label{eq:rpores}
    r=(1-p_j)/a,
\end{equation}
where $1-p_j$ is the specific area of pores, $a=A/L^2$ is the specific total perimeter of
pores, and $A$ is the total perimeter of pores.

The same as for jamming concentration $p_j$, the values of $r$ are practically equivalent at $s=0$ and $s=1$ for both RSA and RRSA models.
Fig.~\ref{fig:porek} compares the effective pore radius $r$ \emph{vs.} value of $k$ (a) and jamming concentration $p_j=p_j(L \to \infty)$ (b) at $s=0$ and $s=1$ (RSA and RRSA models). The value of $r$ increases with $k$ and $s$ values growth (Fig.~\ref{fig:porek}(a)). Moreover, direct correlations between the pore radius $r$ and the jamming concentration $p_j$ are observed. Increase of the jamming concentration evidently reflects decrease of the effective pore radius $r$ (Fig.~\ref{fig:porek}(b)).

Interesting correlations are also observed between the effective pore radius $r$ and the order parameter $s$ (Fig.~\ref{fig:pores}). In close analogy with behavior of the jamming concentration $p_j(s)$, dependencies $r$ were rather different for short ($k<4$) and longer $k$-mers. If $k$-mers are short, increase of $s$ results in increase of the effective pore radius $r$. The opposite behavior is observed for longer $k$-mers  ($k>8$). At same values of $s$ and $k$, the RSA model demonstrates better jamming packing and smaller effective pore radius.

\subsubsection{Jamming concentration vs. length of $k$-mer}

For 1d problem, the jamming concentration \emph{vs.} length of $k$-mer can be evaluated exactly as expressed by Eq.~\ref{eq:ue1Djamm}.
Surprisingly enough, this very complex $p_{j}(k)$ dependence may be fitted rather well by a simple power law (Eq.~\ref{eq:power}) with $p_{j}(\infty)=c_{R}$~\cite{Renyi1958}, $a= 0.2277\pm 0.0019$, $\alpha =1.0111 \pm 0.0023$, $\rho=0.99997$ (coefficient of determination).

\begin{figure}[h] 
\centering
\includegraphics[width=\linewidth]{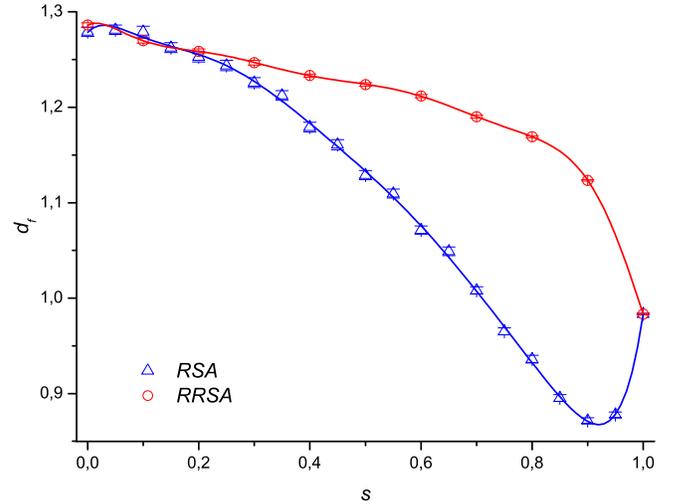}
\caption {Fractal dimension of the RSA and RRSA jamming networks $d_f$  \emph{vs.} order parameter $s$.} \label{fig:df}
\end{figure}

Fig.~\ref{fig:Pvsk_pow} shows jamming concentration $p_j$ \emph{vs.} length of linear segment $k$ at different anisotropy $s$ for RSA and RRSA models). It has been found that numerical results may be rather well fitted by a power law function Eq.~\ref{eq:power}. However, for problem of nonrandomly oriented linear $k$-mers, $p_{j}(\infty)$, $a$ and, $\alpha$ depend on the order parameter $s$ (Table~\ref{tab:pjvsSkoo}). For instance, for isotropic problem ($s=0$) the obtained data (Table~\ref{tab:pjvsSkoo}) are in satisfactory agreement with the previously reported data: $p_{j}(\infty)=0.66\pm 0.01$, $a \approx 0.44$ and $\alpha \approx 0.77$~\cite{KondratPre63}. The fitting parameters obtained for completely ordered $k$-mers, ($s=1$) (Table~\ref{tab:pjvsSkoo}) are also in satisfactory correspondence with the theoretical predictions for 1d problem.

We should emphasize that for the studied problem of nonrandomly oriented $k$-mers, the jamming concentration decreases as an inverse power of the linear segment length, \emph{i.e.}, $p_j \propto k^{-\alpha}$ and the exponent $\alpha$ is not universal. Note that power law behavior with exponent $\alpha\approx 0.2$ was observed for off-lattice RSA of rectangles~\cite{Ziff1990JPhysA} and it was suggested that indicates about fractal structure of the jamming networks with a dimension $d_f=2-\alpha\approx 0.2$. For completely ordered $k$-mers, ($s=1$) the value of $d_f$ is close to $1$ and it is expected for 1d problem. For disordered systems at $s<1$ the fractal dimension $d_f$ increases for RRSA model and goes through  minimum for RSA model when the order parameter $s$ decreases (Fig.~\ref{fig:df}).

\section{\label{sec:concl}Conclusion}
Two different models describing jamming of the partially ordered linear $k$-mers ($k=2$--$128$) with predetermined order parameter $s$ on a square lattice are discussed. In usual RSA model, the substrate "selects" the $k$-mer with appropriate orientation, which results in amplification of the actual order parameter, $s_0>s$. In relaxation RSA model (RRSA), the order parameter remains at the predetermined level, $s_0\approx s$. The similar jamming behavior is observed at different values of $k$ for both RSA and RRSA models in problems of completely disordered (at $s=0$) and ordered (at $s=1$) systems. Our new simulation results for jamming concentrations $p_j$ are in excellent agreement with simulation data ($s=0$, $k=2,3,4,8$)~\cite{Evans,Vandewalle,KondratPre63,KondratPre64,Cherkasova} and analytical results~\cite{Krapivsky2010} previously published for the problem of completely disordered system. For short $k$-mers ($k\leqslant 4$), more dense configurations are observed in disordered systems ($s=0$) as compared with those completely ordered ($s=1$). However, an opposite behavior is observed for longer $k$-mers.
In partially oriented systems ($s<1$), the jamming configurations consist of the blocks of vertically and horizontally oriented $k$-mers ($v$-and $h$-blocks, respectively), and large voids between them. The $v$-blocks merge into the percolation cluster when the order parameter $s$ exceeds some critical value $s_c$, which depends upon $k$ and model of deposition.  The critical exponent of random percolation $\nu=4/3$ is observed in the scaling relations. It is demonstrated that in the intermediate region of $0<s<1$ the RSA process allows obtaining of better jamming packing and smaller effective pore radius than RRSA process at the same values of $s$ and $k$.  For RSA model, the value of $p_j(k)$ monotonically decreases ($k\leqslant 4$) or increases ($k>4$) when the order parameter $s$ increases. For RRSA model, the value of $p_j(k)$ goes through the minimum at $k\geqslant 3$ when $s$ increases. Finally, it is found that numerical results for $p_j$ \emph{vs.} $k$ dependence may be rather well fitted by a simple power law function $p_j \propto k^{-\alpha}$ with parameters dependent on the order parameter $s$. The power exponent $\alpha$ grows when the order parameter $s$ increases from $\alpha\approx0.72$ at $s=0$ to $\alpha\approx 1$ at $s=1$. The jamming networks display fractal properties with fractal dimension $d_f\approx 1$ for completely ordered $k$-mers, ($s=1$) and value of $d_f$ increases for RRSA model and goes through  minimum for RSA model when the order parameter $s$ decreases (Fig.~\ref{fig:df}).

\begin{acknowledgments}
Work is partially supported by the Russian Foundation for Basic
Research, grants no.~09-02-90440-Ukr\_f\_a, 09-08-00822-a, and
09-01-97007-r\_povolzhje\_a.
\end{acknowledgments}

\end{document}